\documentclass[manuscript]{aastex}
\usepackage{graphicx}
\usepackage{epstopdf}

\shorttitle{Emission Measures and Electron Temperatures for Supernova Remnants}
\shortauthors{Leahy}

\begin{document}


\title{Emission Measures and Emission-measure-weighted Temperatures of Shocked ISM and Ejecta in Supernova Remnants}


%\correspondingauthor{Denis Leahy}
%\email{leahy@ucalgary.ca}

\author{Denis Leahy, Yuyang Wang, Bryson Lawton, Sujith Ranasinghe}
\affil{Department of Physics $\&$ Astronomy, University of Calgary, Calgary, Alberta T2N 1N4, Canada}
%\author{Yuyang Wang}
%\affil{Department of Physics $\&$ Astronomy, University of Calgary, Calgary,Alberta T2N 1N4, Canada}
%\author{Bryson Lawton}
%\affil{Department of Physics $\&$ Astronomy, University of Calgary, Calgary,Alberta T2N 1N4, Canada}
\and
\author{Miroslav Filipovi\'c}
\affil{Western Sydney University, Locked Bag 1797, Penrith South DC, NSW 1797, Australia}
%\author{Bryson Lawton}
%\affil{Department of Physics $\&$ Astronomy, University of Calgary, Calgary, Alberta T2N 1N4, Canada}


\begin{abstract}
{A goal of supernova remnant (SNR) evolution models is to relate fundamental parameters of a supernova (SN) explosion and 
progenitor star to the current state of its SNR.  
The SNR hot plasma is characterized by its observed X-ray spectrum, which yields electron temperature, emission measure and abundances.
Depending on their brightness, the properties of the plasmas heated by the SNR forward shock, reverse shock or both can be measured. 
The current work utilizes models which are spherically symmetric.
One dimensional
hydrodynamic simulations are carried out for SNR evolution prior to onset of radiative losses.
From these, we derive dimensionless emission measures and emission-measure-weighted temperatures, and 
we present fitting formulae for these quantities as functions of scaled SNR time. 
These models allow one to infer SNR explosion energy, circumstellar medium density, age, ejecta mass and ejecta density profile 
from SNR observations. The new results are incorporated into the SNR modelling code SNRPy.
The code is demonstrated with application to three historical SNRs: Kepler, Tycho and SN1006.
}

\end{abstract}




\section{Introduction}

Supernovae (SNe) and supernova remnants (SNRs) have a great impact on the evolution of galaxies and the interstellar medium (ISM) within
galaxies (\citealt{2012Vink} and references therein). 
They do this via their energy input  into the ISM, and return of elements. 
The energy content of SNRs is in the forms of: i) the kinetic energy of the unshocked ejecta, shocked ejecta and shocked ISM; and 
ii) the thermal energy in  shocked ejecta and shocked ISM with temperatures $\sim$1 keV.
The shocked gas is observed in X-rays. 
The synchrotron emission from relativistic electrons accelerated by the SNR shockwave is observed in the radio band. 
%To characterize the evolutionary state of an SNR, X-ray observations are required.
There are $\sim$300 observed SNRs in our Galaxy 
\citep{2014Green}, but only a small number of the  have been well characterized. 
The vast majority don't have well-established inferred supernova (SN) type, explosion energy, age or ISM density. 
Lack of better observational data, including lack of information on the ejecta, is a major obstacle to understanding SNRs. 
Another important
obstacle to characterizing SNRs is the lack of easily-applied models that are still realistic enough to describe SNR evolution.

In order to expedite characterization of a significant number of SNRs, \citet{2017LW} presented a set of SNR evolution models 
and a software implementation in Python, called SNRPy. 
That model included forward and reverse shock radius and velocity for all stages of evolution.
For modelling observations, emission measures ($EM$) and $EM$-weighted temperatures are required. 
The above model included $EM$ and $EM$-weighted temperatures for  
the self-similar early ejecta-dominated (ED) phase (cases (s,n)= (0,7), (0,12) and (2,7)) 
and self-similar Sedov-Taylor (ST) phase.
Even with that limitation, \citep{2017LeahyLMC} demonstrated that model fitting of SNRs can give valuable information on the nature of SN explosions. 
\citet{2018Leahy15GalSNR} used the \citet{2017LW} model but with all self-similar ED cases included (s= 0 and 2, n= 6 through 14).
In order to model the important ED to ST transition time, that work used a linear interpolation of $EM$ and $EM$-weighted temperatures vs. time 
between the end of the ED phase and start of the ST phase.
The fitting of evolution models to 15 Galactic SNRs showed that Galactic SNRs have the same broad range of explosion energies as 
LMC SNRs, but occur in significantly denser ISM \citep{2018Leahy15GalSNR}.
SNR and SN population properties from models are important inputs to understand the evolution of the Galaxy 
and its interstellar medium, e.g. \citet{2001Ferri}. 
%{2017LeahyLMC} {2017LW}{2018Leahy15GalSNR}
%{2017Rana2SNRmol} {2018rana21SNR} {2018Rana3SNRmol}{2018Rana4SNR}

For a typical SNR, with explosion energy $10^{51}$ erg, ejected mass of 1.4 $M_{\odot}$ and ISM density of 1 cm$^{-3}$, 
the ED phase lasts from explosion to 220 yr, the ED to ST transition lasts from 220 yr to 1500 yr, and the ST phase from
1500 yr to onset of radiative cooling at $\sim$15,000 yr. 
Most X-ray detected Galactic SNRs (e.g.  \citealt{2018Leahy15GalSNR}) are either in the ED to ST or the ST phase.
Thus quantitative calculations of $EM$s and $EM$-weighted temperatures are needed in order to model X-ray emission from SNRs.
A comprehensive and easy-to-use set of $EM$s and $EM$-weighted temperatures have not been presented before, 
even for the case of spherically symmetric SNR evolution.
As late as the ST phase, the shocked ejecta continue to emit X-rays and can be an important diagnostic of the SNR and SN
explosion, so calculation of $EM$ and $EM$-weighted temperature is important for both ED to ST and ST phases.

This work focuses on calculating the $EM$s and $EM$-weighted temperatures for the full evolution of an
SNR up to the onset of the radiative phase.
Section 2 of this paper presents an brief overview of SNR evolution, including the unified SNR evolution introduced by 
\citet{1999truelove}. Because of the existence of unified SNR evolution, dimensionless $EM$s and $EM$-weighted 
temperatures, if available from models for both forward-shocked gas and reverse-shocked gas, serve as a 
powerful diagnostic tool for the state of an SNR (\citealt{2018Leahy15GalSNR} and references therein).
%, required for modelling X-ray observations of SNRs. 
%Section 3 describes calculations of interior structure for the self-similar Ejecta-Dominated (ED) phase, and the self-similar ST phase for uniform ISM and cloudy ISM cases.   
Section 3 describes the hydrodynamic simulations which include the early ED phase and the late ST phase,
and the transition between them (ED to ST). 
%Section 4.1 discusses results from the self-similar solutions, including emission measures ($EM$s) and $EM$-weighted temperatures.
Section 4 describes the results from the hydrodynamic simulations, and analytic fits to dimensionless $EM$s and temperatures.
Section 5 gives a summary and where to access the results and the updated SNR modelling code SNRPy. 
In the appendix we present calculations of $EM$ and $EM$-weighted temperatures for the self-similar ED, ST and cloudy ISM cases.

\section{Supernova Remnant Evolution and Structure}

A SN explosion creates a SNR starting with the ejection of the SN progenitor envelope at high speed, typically $\sim$10000 km/s,
with inner envelope moving more slowly and outer envelope moving more quickly.
General descriptions of SNR evolution are given in numerous places (e.g., \citealt{1988cioffi}, \citealt{1999truelove}-  hereafter TM99,
and \citealt{2017LW}- hereafter LW17).
The ejecta collides with the circumstellar medium (CSM) or ISM, causing a foward shock (FS) to propagate outward and a 
reverse shock (RS) to propagate back into the ejecta.

The general sequence of SNR evolution starts with the ED phase for which the effect of the ejected mass is important. 
This gradually evolves to the ST phase, for which the swept-up mass by the SN shock far exceeds the ejected mass. 
For ED, transition and ST phases, radiative energy losses are unimportant. 
Beyond the ST phase, radiative losses become important (e.g. \citealt{1988cioffi}). 
In the current work, the phases prior to the radiative phases are considered.

The basic interior structure of a SNR, prior to the ST phase, has the following regions from outside to inside:
i) the undisturbed CSM;
ii) the FS moving into the CSM; a layer of shocked CSM; 
iii) the contact discontinuity (CD) separating the shocked CSM from the shocked ejecta;
iv) the layer of shocked ejecta;
v) the RS moving inward relative to the ejecta; and
vi) the undisturbed ejecta.
The unshocked ejecta has a homologous velocity profile ($v \propto r$ at fixed time). 

After the reverse shock reaches the center of the SNR, the entire ejecta is fully shocked.
Reflected shocks and sound waves are generated at this time, and die out slowly over time \citep{1988cioffi}.
The reflected shocks and sound waves are clearly seen in the numerical simulations presented below.

In order to calculate SNR evolution and structure, the following simplifying assumptions are made. 
The SNR is spherically symmetric and radiative losses have not yet set in.
This allows us to use the unified SNR evolution model of \citep{1999truelove}, where powerful scaling relations
apply. This means that a single set of hydrodynamic calculations can be made, and the results applied to 
explosions with different explosion energy, different ejecta mass, and different ISM density by 
using scaling relations.
The CSM has: i) constant density; or ii) $1/r^2$ stellar wind density profile centered on the SN, 
i.e., $\rho_{CSM}=\rho_s r^{-s}$ with s=0 or s=2.
The unshocked ejecta has a constant density core for $r\le R_{core}$, and a power-law density envelope: 
$\rho_{ej}\propto r^{-n}$ for $r>R_{core}$. 

%With these assumptions, the forward and reverse shock radii were calculated vs. time for ED through ST phases by TM99. 
%The early part of the ED phase follows a self-similar evolution \citep{1982Chev}.
%Because there is no smooth transition for s=2 from ED to ST phases (e.g. TM99) only the ED phase is calculated here for the s=2 case.
%The self-similar evolution of a SNR in a cloudy ISM \citep{1991WL} (hereafter  WL91) is also calculated here.
 %The number density of electrons is $n_e$, of hydrogen nuclei is $n_{H}$, and of  total number of ions is $n_{ion}$. 
%Then the total mass density is given by $\rho=\mu_e n_e m_{H}= \mu_{H} n_{H} m_{H}= \mu_{ion} n_{ion} m_{H}=\mu_{tot} n_{tot} m_{H}$.
%Here $\mu_e$ is the mean mass per electron, $\mu_{H}$ is the mean mass per hydrogen nucleus, $\mu_{ion}$ is the mean mass per ion and $\mu_{tot}$ is the mean mass per particle (with  $n_{tot}=n_e+n_{ion}$ is the total number of particles).
 
\subsection{Characteristic Scales}

Non-radiative supernova remnants undergo a unified evolution (TM99). 
The characteristic radius and time for $s=0$ are given by
$R_{ch}=(M_{ej}/\rho_0)^{1/3}$ and $t_{ch}=E_0^{-1/2}M_{ej}^{5/6}\rho_0^{-1/3}$ 
with $M_{ej}$ the ejected mass and $E_0$ the explosion energy.
%$\rho_0$ is the ISM mass density, $\rho_0=\mu_{H} n_{0} m_{H}$, where $n_0 is the ISM hydrogen number density is $
The characteristic velocity is $V_{ch}=R_{ch}/t_{ch}$ and characteristic
shock temperature is $T_{ch}=\frac{3}{16}\mu \frac{m_H}{k_B}V_{ch}^2$, with  $\mu$  the mean mass per particle.
For SNR in a CSM with $s=2$, the  characteristic radius and time are given by
$R_{ch}=(M_{ej}/\rho_s)$ and $t_{ch}=E_0^{-1/2}M_{ej}^{3/2}v_w/\dot{M}$,
with $\dot{M}$ and $v_w$ the wind mass loss rate and velocity, and $\rho_s=\frac{\dot{M}}{4\pi v_w}$.

\subsection{Emission Measure ($EM$), $EM$-weighted Temperature and Column $EM$}

Because the emission from the hot shocked gas in a SNR is dominated by two body processes, it
depends on the product of electron and ion densities (e.g. \citealt{1976Ray}).
$EM$ is defined in terms of electron density $n_e$ and hydrogen ion density $n_H$ by $EM=\int n_e(r) n_H(r) dV$.  
%EM is usually expressed in units of  $10^{58}$cm$^{-3}$.
% $n_H$ and the ion density $n_{ion}$ are related by the mean molecular weights.
$EM$ can be measured by X-ray observations, so the measured $EM$ is critical to determining the evolution state of a SNR.
 
$EM$ can be calculated from a model SNR density profile. During self-similar phases of a SNR, the density profile has a constant 
functional form with normalization and scaling with radius dependent on time.
The dimensionless $EM$, $dEM$, was defined by LW17 as $dEM=EM/(n_{e,s}n_{H,s}R_{FS}^3$)
with $n_{e,s}$ and $n_{H,s}$ are $n_e$ and  $n_H$ immediately inside the forward shock (FS).
We extend this to define $dEM_{FS}$ and $dEM_{RS}$ for the gas heated by the
FS and for gas heated by the RS, respectively. 
\begin{eqnarray}
dEM_{FS} &  = & \frac{1}{n_{e,s}n_{H,s}R_{FS}^3}\int_{R_{CD}}^{R_{FS}}  n_e(r) n_H(r) dV \\
dEM_{RS} &  = & \frac{1}{n_{e,s}n_{H,s}R_{FS}^3}\int_{R_{RS}}^{R_{CD}}  n_e(r) n_H(r) dV
\end{eqnarray}

The observed temperature of a SNR, derived from the X-ray spectrum, 
depends on the state of the SNR and on the adopted X-ray spectrum model. 
Commonly used models are one or two component non-equilibrium ionization models.
For example, see \citealt{2016Maggi} for LMC SNR fits and \citealt{2018Leahy15GalSNR} for a summary of models for 
15 non-historical Galactic SNRs. 
In some cases, a SNR has an observed multiple-electron temperature plasma which cannot be attributed 
to the expected two single electron temperature plasmas from forward ISM and reverse shocked ejecta, respectively.
ln this case, the plasma is more complicated than the model assumes, and further approximations are required. 
If there is a dominant electron temperature, the model
can be applied assuming the forward or reverse shocked ejecta is dominated by that temperature.
If there is not a dominant electron-temperature component, another possible approximation is to use the mean observed electron 
temperature.

For a single component plasma emission model, 
the X-ray temperature measures the $EM$-weighted temperature of that component of shocked-heated gas.
A strong shock with speed $V_s$ heats the ions to the ion shock temperature $k T_{ion}=\frac{3}{16}\mu_{ion} m_{H} V_s^2$.
We use the formula for electron heating by Coulomb collisions described in section 3.1.3 (equation 1) of 
LW17, which uses the formulation of \citet{1982CA}. 
This gives a steady increase of the ratio of electron temperature to ion temperature from 
$T_e/T_{ion}$=0.03 at age 0 yr to $T_e/T_{ion}$=1 at age of $\sim$1000-5000 yr. 
The latter number depends on CSM density, ejecta mass, and explosion energy. 
 LW17 defined the dimensionless temperature $dT=\frac{1}{T_{FS}}\frac{1}{EM}\int  n_e(r) n_H(r) T(r) dV$, 
 with $T_{FS}$ the forward shock temperature.
We extend this to define $dT_{FS}$ and $dT_{RS}$ for the two plasma components heated by the
FS and by the RS, respectively. 
\begin{eqnarray}
dT_{FS} &  = & \frac{1}{T_{FS}}\frac{1}{EM}\int_{R_{CD}}^{R_{FS}}  n_e(r) n_H(r) T(r) dV \\
dT_{RS} &  = & \frac{1}{T_{FS}}\frac{1}{EM}\int_{R_{RS}}^{R_{CD}}  n_e(r) n_H(r) T(r) dV
\end{eqnarray}

The surface brightness of a SNR depends on the line-of-sight integral of the emission coefficient 
$j(\nu)=  n_e n_H \epsilon(\nu)$, with emissivity $\epsilon(\nu)$.
The column emission measure ($CEM$) is often used as a proxy for surface brightness, valid when the emission coefficient is 
only weakly dependent on the temperature history of the parcel of gas (e.g. see \citealt{1991WL}, herafter WL91).
$CEM$ is given by $CEM(B)=\int  n_e n_H dS$, where the integral is along the line of sight through the SNR 
at impact parameter $B$ from center. 

We define the dimensionless $CEM$  using the scaled densities and dimensionless impact parameter, $b=B/R_{FS}$, by 
\begin{equation}
cEM(b)=\int  \frac{n_e(x(s))}{n_{e,s}} \frac{ n_H(x(s))}{n_{H,s}} ds 
\end{equation}
with  $s=S/R_{FS}$ and $x(s)=\sqrt{b^2+s^2}$.
More generally, we define the dimensionless $cEM(b)$ separately for gas heated by the forward shock and 
gas heated by the reverse shock, dimensionless $cEM_{FS}(b)$ and $cEM_{RS}(b)$. For  $cEM_{FS}(b)$, $x(s)$
is limited to values between $R_{CD}/R_{FS}$ and 1, while for $cEM_{RS}(b)$, $x(s)$
is limited to values between $R_{RS}/R_{FS}$ and $R_{CD}/R_{FS}$.

 \section{Hydrodynamic Calculations of SNR Structure}
 
To calculate the interior structure of a SNR, we use the hydrodynamic equations.
The evolution follows a unified evolution as shown by TM99, before radiative losses become important. 
Unified evolution means that solutions have the same dependence on $t/t_{ch}$ if %density is scaled by $\rho_{ch}$ 
radius is scaled by $R_{ch}$, velocity is scaled by $V_{ch}$ and temperature is scaled by $T_{ch}$.
%For s=2 in an infinitely extended wind, the solution remains self-similar for all time.
Because there is no smooth transition for s=2 from ED to post-ED (e.g. TM99), 
we calculate the post-ED phases only for the s=0 case.
That evolution is the subject of this section.

The evolution of $R_{FS}$ and $R_{RS}$ has been calculated using an analytic approximation for ED, ED to ST and ST phases by TM99. 
%They consider the full non-radiative evolution of a SNR, which ends with the start of transition to the PDS radiative phase at time $t_{pds}$. 
The reverse shock slows its outward motion relative to the ISM at about the
time that it reaches the ejecta core, at time $t_{core}$ (TM99).
Then it propagates inward, reaching the center of the SNR at time $t_{rev}$ (TM99).
The evolution is continuous, but it is useful to label the phases as `ED' for $0<t<t_{core}$, 
`ED to ST' for  $t_{core}<t<t_{rev}$, and `ST' for $t_{rev}<t<t_{PDS}$, where $t_{PDS}$
is the time where radiative losses affect the evolution (\citealt{1988cioffi}, TM99 and LW17).
However, the `ST' phase can be quite different that the `pure ST' evolution, as pointed out by LW17.
 
Here we calculate the evolution and interior structure for s=0 and n=6 to 14 using the hydrodynamics code PLUTO
(\citealt{2007mignone}, \citealt{2012mignone}).
A core-envelope structure for the ejecta is assumed. %, with constant density core and power-law density envelope of index n. 
%For $n>5$ the outermost velocity of the envelope $v_{ej}$ does not affect the mass or energy of the envelope so that the parameter
%$v_{core}/v_{ej}$ can be taken to be zero (by taking the limit $v_{ej}=\infty$).
For the simulations the fundamental code units were set to $\rho_u=10^{-18}$ gm/cm$^{-3}$ (density), $r_u=10^{16}$ cm (distance)
and  $v_u=10^{7}$ cm/s (velocity).
The resulting code units for time, pressure, mass and energy are $t_u=10^{9}$ s, $P_u=10^{-4}$ dyne cm$^{-2}$ ,  $M_u=10^{30}$ gm and
$E_u=10^{44}$ erg.

We tested different values for the ISM density, ejecta mass and explosion energy to verify that SNR evolution in scaled 
variables (density scaled by $\rho_{ISM}$, time scaled by $t_{ch}$, radius by $R_{ch}$, velocity by $V_{ch}$
and pressure by $P_{ch}=\rho_{ISM}V_{ch}^2$) was independent of those initial quantites.
Then we set the ISM density $=10^{-22}$gm/cm$^{3}$, ejected mass $=1~M_{\odot}$, and explosion energy
$=10^{51}$ erg for the remaining calculations.
This yields characteristic scales of  $t_{ch}=3.839\times10^{9}$ s $=121.7$ yr,
$R_{ch}=2.714\times10^{18}$ cm $=0.8797$ pc, $V_{ch}=7.071\times10^{3}$ km/s
and $P_{ch}=5.000\times10^{-5}$ dyne cm$^{-2}$. 

The SNR initial conditions were initially taken to be an unshocked ejecta with core and envelope components, 
plus shocked and unshocked ISM.
The resulting time-dependent solutions showed large transient fluctuations in the hydrodynamic variables. 
%as the initial conditions evolve into a consistent solution of the equations. 
Use of more accurate initial conditions should result in smaller transient fluctuations.
Thus a more accurate second case of initial conditions was constructed from the self-similar 
Chevalier-Parker (CP, \citealt{1982Chev}, \citealt{1963Parker})
solutions, consisting of unshocked and shocked ejecta, and shocked and unshocked ISM. 

For the first case initial conditions, a small outer ejecta radius $R_{ej}=5\times10^{12}$ cm was chosen. 
The core radius was taken as $10^{-1.5}$ of $R_{ej}$. The core density was set so that the integrated mass
from $r=0$ to $r=R_{ej}$ was 1 $M_{\odot}$. The velocity increases linearly with radius for the unshocked ejecta,
so the velocity profile is specified by the velocity $v_{ej}$ at $r=R_{ej}$.  $v_{ej}$ was determined by requiring the total
ejecta kinetic energy to be the explosion energy. % integrated from density and velocity profiles
For example, $v_{ej}=2.888\times10^{5}$ km/s and $v_{core}=9.132\times10^{3}$ km/s is found for n=7. %vej=40.839*7.071e8
%The ejecta density and velocity profiles are then fully specified. 
The ejecta pressure was set to a low value ($10^{-8}$ dyne cm$^{-2}$).

\begin{figure}
%\plottwo{s0n7_init.png}{s0n7_initCP.png}
\includegraphics[angle=0,scale=0.55] {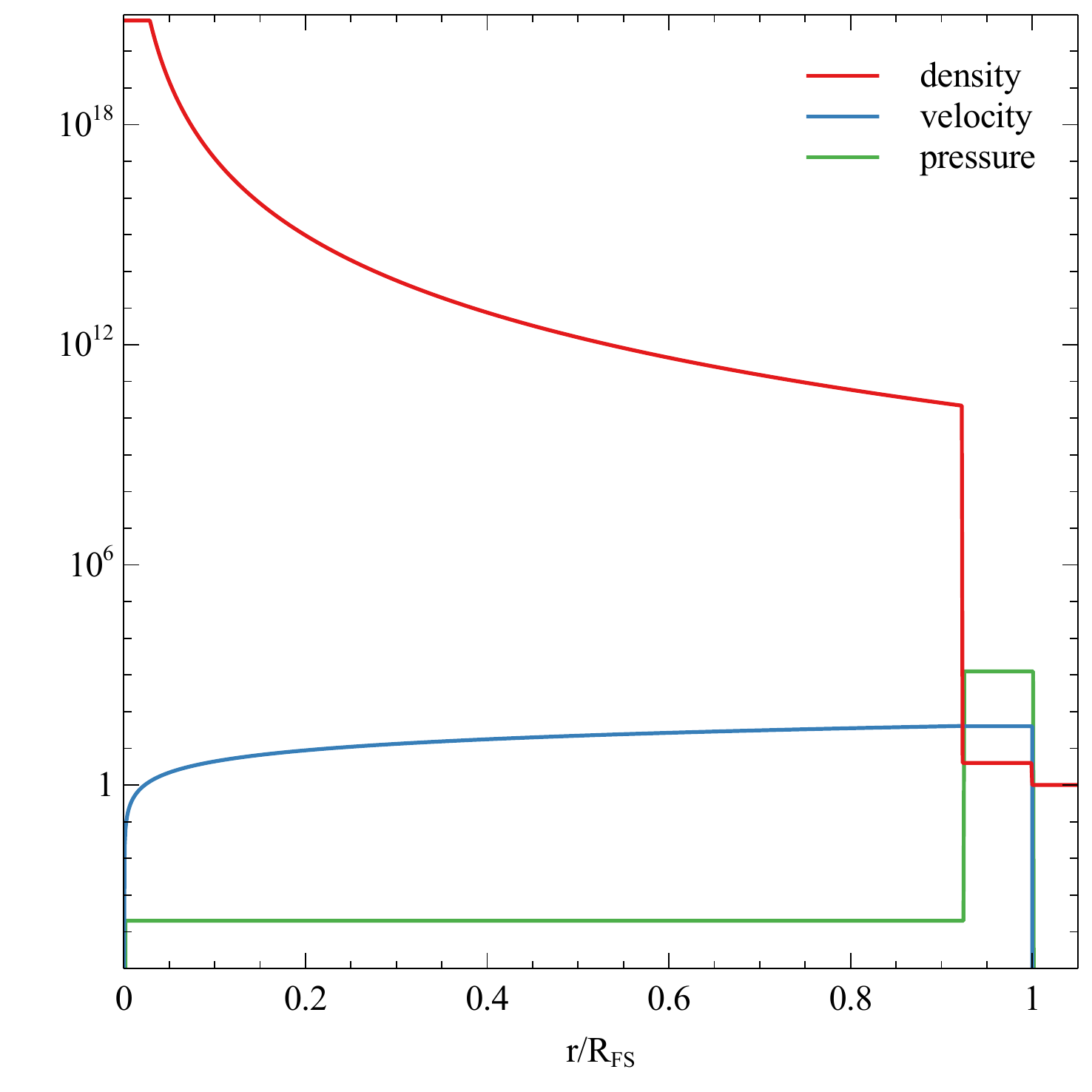}
\includegraphics[angle=0,scale=0.55] {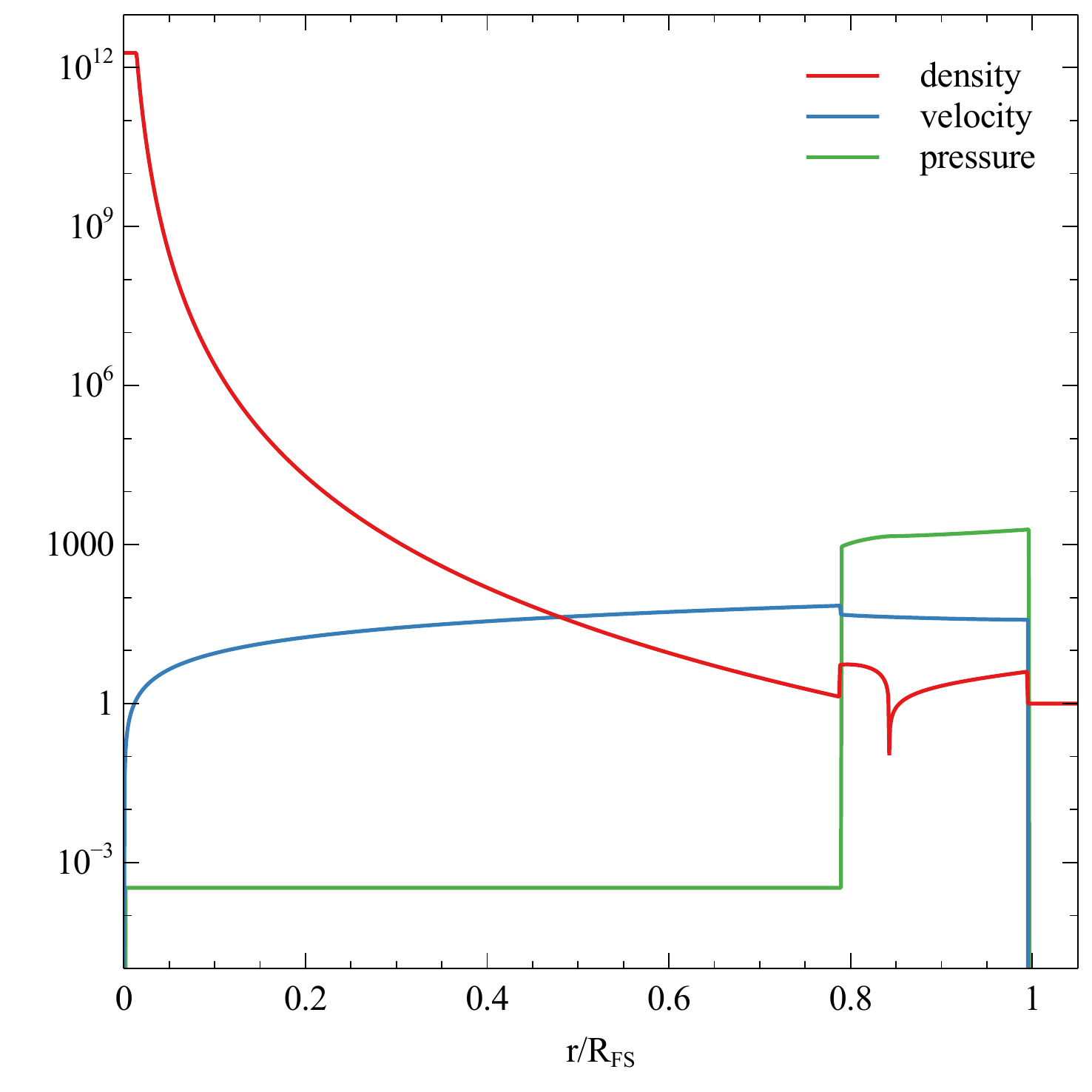}
\caption{Simulation of an SNR with s=0, n=7:
Left panel: initial conditions of unshocked ejecta and shocked ISM at t=$4.5\times10^{-8}$, $R_{FS}$=$2.0\times10^{-6}$.
Right panel: initial conditions using CP self-similar solution of shocked ejecta and shocked ISM
 at t=$4.1\times10^{-5}$, $R_{FS}$=$3.3\times10^{-3}$.
Density, velocity, pressure, time t and forward shock radius $R_{FS}$ are in characteristic units, the x-axis is in units of $r/R_{FS}$.
 \label{figinit7}}
\end{figure}

A layer of shocked ISM with density 4 $\rho_{ISM}$ was added outside the ejecta from $R_{ej}$ to $R_{FS,0}$.
The outer initial forward shock radius is $R_{FS,0}=(13/12)R_{ej}$, determined from the 
requirement that the mass of shocked ISM equals the total ISM mass swept up between $r=0$ and $R_{FS,0}$.
The velocity of the layer is $v_{ej}$, because the high density at early times of the ejecta makes
it act like a rigid piston. This yields a shock velocity at $R_{FS,0}$ of $v_{FS,0}=(4/3)v_{ej}$ and an interior pressure of
the ejecta layer of $(3/4)\rho_{ISM}v_{FS,0}^2$.
%The explosion energy was set to $10^{51}$ erg.
Including the time for the outer edge of the ejecta to expand to $R_{ej}$ from $r=0$, % at velocity $v_{ej}$,
the initial solution has $t/t_{ch}=4.5\times10^{-8}$ and $R_{FS}/R_{ch}=1.99\times10^{-6}$ for n=7.
The initial conditions for the n=7 simulation are shown in the left panel of Fig.~\ref{figinit7}.

The second case for the initial conditions utilizes the CP self-similar solutions. %, computed in Section~\ref{sec:ED} above.
%and thus are numerical.
%The self-similar solutions for n>5 have an outer edge of the unshocked ejecta at the reverse shock, $R_{RS}$. 
The initial density profile is determined by matching both the unshocked ISM and the unshocked ejecta to the CP solution. 
The outer boundary of the CP solution at $R_{FS,0}$ is set to $4\rho_{ISM}$, and the outer boundary of the unshocked ejecta 
at $R_{RS,0}$ is set to $(1/4)\rho_{CP}(R_{RS,0})$ from the CP solution.
%The core size and the ejecta velocity are adjusted to obtain the desired ejecta mass and total energy, which are determined
%by integration over the unshocked ejecta and CP solutions.

The ejecta mass includes the core, $r<R_{core}$, the unshocked powerlaw envelope, $R_{core}$ to $R_{RS}$, 
and the shocked ejecta, $R_{RS}$ to $R_{CD}$.
The shocked ejecta contains a pileup of the shocked envelope mass.% which originally extended from $R_{RS}$ to $\infty$ with powerlaw density profile. 
We define $w_{core}=R_{core}/R_{RS}$, which is different than TM99 who don't have a layer of shocked ejecta in their initial density
profile. We set $w_{core}=10^{-1.5}$.
For small $w_{core}$, the contribution from the shocked ejecta is small.
Thus we use the analytic value of mass from the unshocked ejecta, 
$M_{ej}\simeq M_{core}+M_{env}=M_{core}\frac{n-3w_{core}^{n-3}}{n-3}$ to obtain the initial estimate of $M_{core}$ from $M_{ej}$.
We integrate to obtain an accurate value of all three contributions to the ejecta mass. 
Then we set the total to the desired ejecta mass, $1 M_{\odot}$, to determine a more accurate value of $M_{core}$ .
$\rho_{core}$ is found from $M_{core}=(4/3)\pi R_{core}^3  \rho_{core}$ for given $R_{core}$.
$R_{core,0}$ is chosen large enough to be resolved with enough grid cells in the hydrocode. %we chose a minimum of 20
It is small enough to give $R_{FS,0}<<R_{ch}$, to include enough of the early ED phase prior to the post-ED evolution.
E.g., for n=7 we chose $R_{core,0}=4.0\times10^{14}$ cm, yielding $R_{FS,0}=1.60\times10^{16}$ cm $=3.31\times10^{-3}R_{ch}$. 

The initial estimate of $v_{core}$ is obtained from the energy of core and unshocked envelope.
This is given by $E_{ej}\simeq E_{core}+E_{env}=E_{core}\frac{n-5w_{core}^{n-5}}{n-5}$ and %the exact relation
$E_{core}=(2/5)\pi R_{core}^3  \rho_{core}  v_{core}^2$. The error in $v_{core}$ is small for small $w_{core}$.
A  more accurate $v_{core}$ is obtained by integrating the energy in the core, unshocked envelope and
the shocked envelope and setting to the explosion energy, $10^{51}$ erg.
The velocity at the outer edge of the unshocked envelope is $v_{env}= v_{core}\times R_{RS,0}/R_{core,0}$. 
The time since explosion for the initial solution is given by $t_{0}=R_{core,0}/v_{core}$.
For n=7,  $v_{core}=9.13\times10^{3}$ km/s and $t_{0}/t_{ch}=4.14\times10^{-5}$. %t=4.38\times10^5 s

To match velocities with the CP solution, we apply shock jump conditions at both forward and reverse shocks. % including pressure conditions.
The postshock pressure at $R_{FS}$ is $P_{FS}=(3/4)\rho_{ISM}V_{FS}^2$. % with $V_{FS,0}$ a free parameter. 
The pressure ratio $x_{RF}=P_{RS}/P_{FS}$ is an n-dependent constant given by the CP solution.  
The reverse shock velocity, relative to the envelope gas, is $P_{RS}=(3/4)\rho_{env}V_{RS}^2$
with the reverse shock velocity in the envelope frame $V_{RS}=v_{env}-V_{RS,obs}$ with $V_{RS,obs}$ the reverse shock
velocity in the observer frame. The gas velocity relative to the post-shock gas $v_{sh,rel}$
is 1/4  of the pre-shock gas $v_{un,rel}$: $v_{sh,rel}=(1/4)v_{un,rel}$. %This is applied to both forward and reverse shocks. 
After a bit of algebra we find:
\begin{equation}
V_{FS}=\frac{4~v_{env}}{3(\sqrt{y_{RF}~\rho_{ISM}/\rho_{env}}+y_{RF})}
\end{equation}
where the ratio of post-shock gas velocities from the CP solution is given by $y_{RF}=v_{sh,RS}/v_{sh,FS}$.

The above procedure fully determines the initial conditions which satisfy the shock jump conditions at both shocks
and have the correct total energy and ejecta mass.
The initial CP solution for n=7 is shown in the right panel of Fig.~\ref{figinit7}. 
This has $v_{env,0}=2.89\times10^{5}$ km/s and  $V_{FS,0}=2.09\times10^{5}$ km/s.

The initial conditions for case 1  were computed analytically using a modified init.c program in PLUTO. 
% which feeds initial conditions into PLUTO. %which produces the hydrocode output.
The initial conditions for case 2 consist of a binary file which includes the CP numerical solutions matched to the unshocked ejecta and and the ISM.
 %by combining the $r<R_{RS,0}$ solutions described
%above, the ISM values for $r>R_{FS,0}$ and the CP numerical solutions
%is read into PLUTO. % as the initial conditions.
For both cases, we added  a passive scalar tracer field to track the contact discontinuity and the ejecta core-envelope boundary. 
%with better accuracy.  
%included with value 12 for $0<r<r_{core}$, 11 for $r_{core}<r<r_{CD}$ and 11 for $r_{CD}<r$.

The SNR evolution includes a large range in time and spatial scales. The typical initial time  is $\sim10^{-8}t_{ch}$ (case 1) to
 $10^{-4}t_{ch}$ (case 2) and initial radius  is $\sim10^{-6} R_{ch}$ (case 1) to  $10^{-3} R_{ch}$ (case 2).
 For case 1, we started the simulation at very early time and small radius in order to allow the approximate
 initial conditions to relax to a more accurate solution. 
The late stage time is $\sim10^{4} t_{ch}$ and late stage radius  is $\sim10^{2}$ $R_{ch}$, for both cases.
Thus it is not possible to compute the SNR structure in a single run of PLUTO. 
Instead we ran the code successively in stages, with the output of each stage used as input for the next stage. 
The computational grid was chosen so that the SNR initial outer shock radius was $\simeq$1/5 of the grid size which
allowed room for the SNR to expand to the edge of the grid before initiating a new stage.
The spatial grid size was chosen 5000 points, so the SNR was resolved by a minimum of 1000 points at any time.
Typically, 7 to 10 stages were computed for each evolution, allowing a %$5^7$ to $5^{10}$ 
$\simeq10^5~-~10^8$ factor in radial expansion.
The time steps were adjusted by PLUTO to satisfy the Courant condition, yielding $\sim80,000$ timesteps per stage. 
The times for saving structure files (or snapshots) of the evolution were chosen manually, resulting in $\sim250$ snapshots per evolution. 

\section{Results and Discussion}
%\subsection{Hydrodynamic SNR Solutions} 

\begin{figure}
%\plotone{hy_RfRrVf_n8.pdf}
\includegraphics[angle=0,scale=0.55] {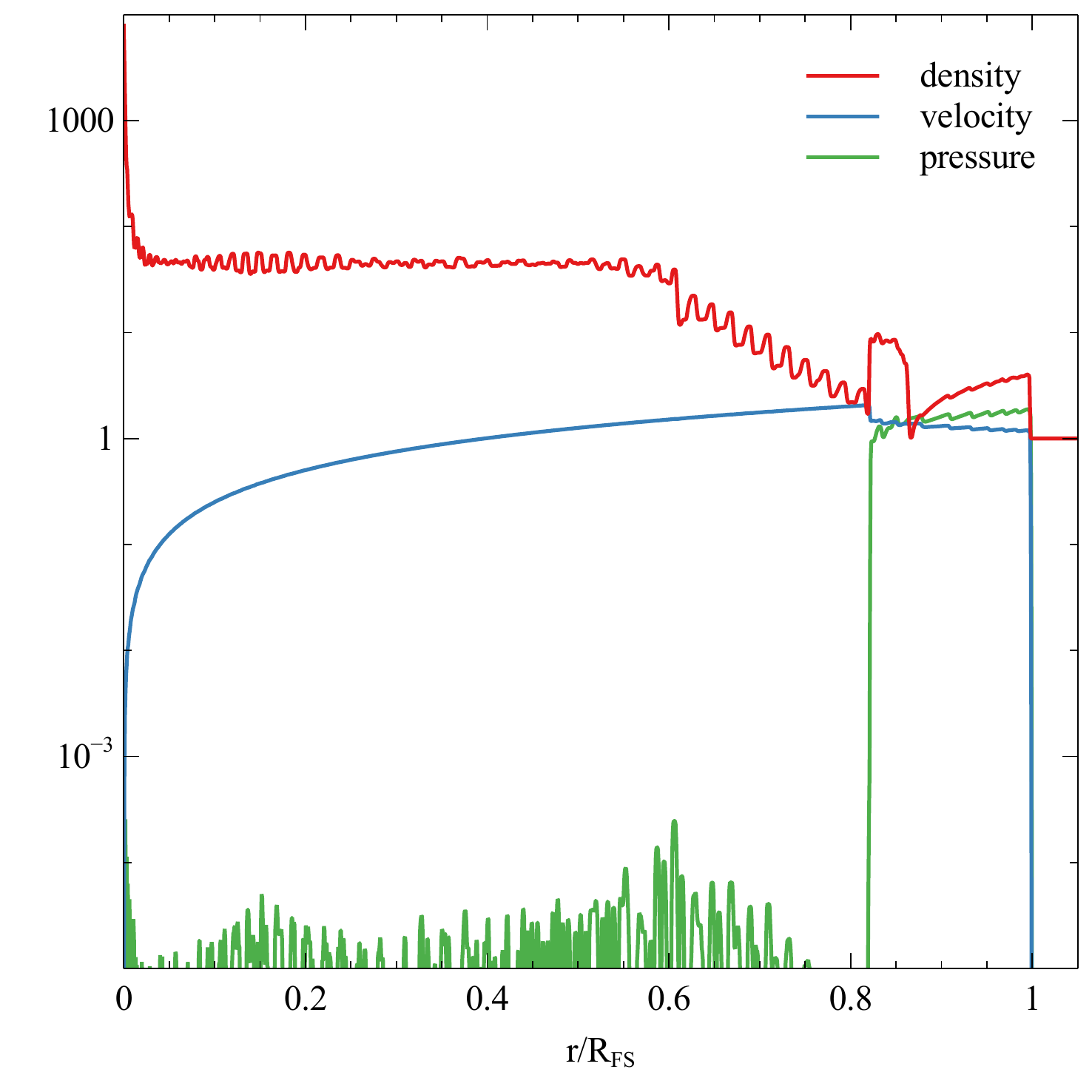}
\includegraphics[angle=0,scale=0.55] {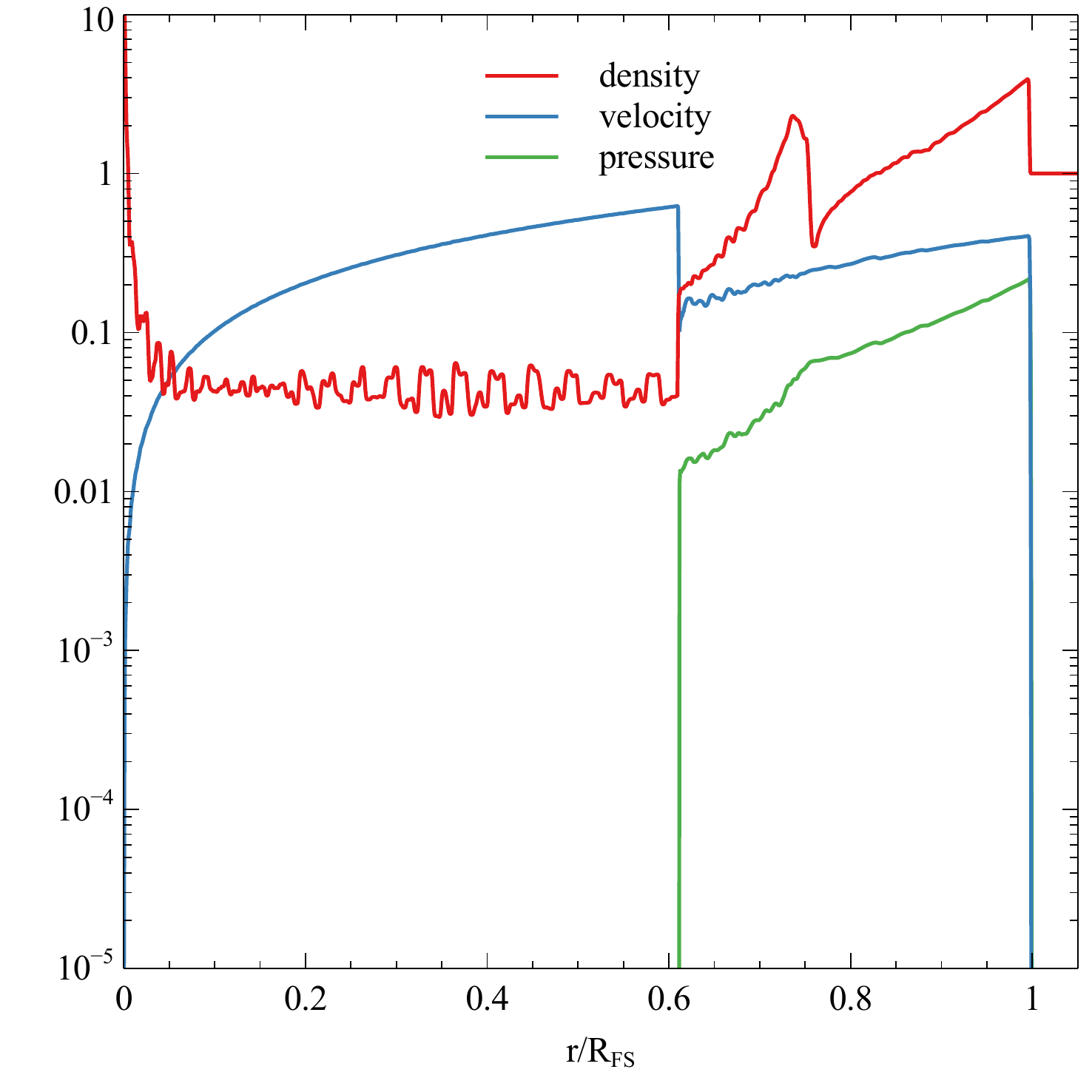}
\includegraphics[angle=0,scale=0.55] {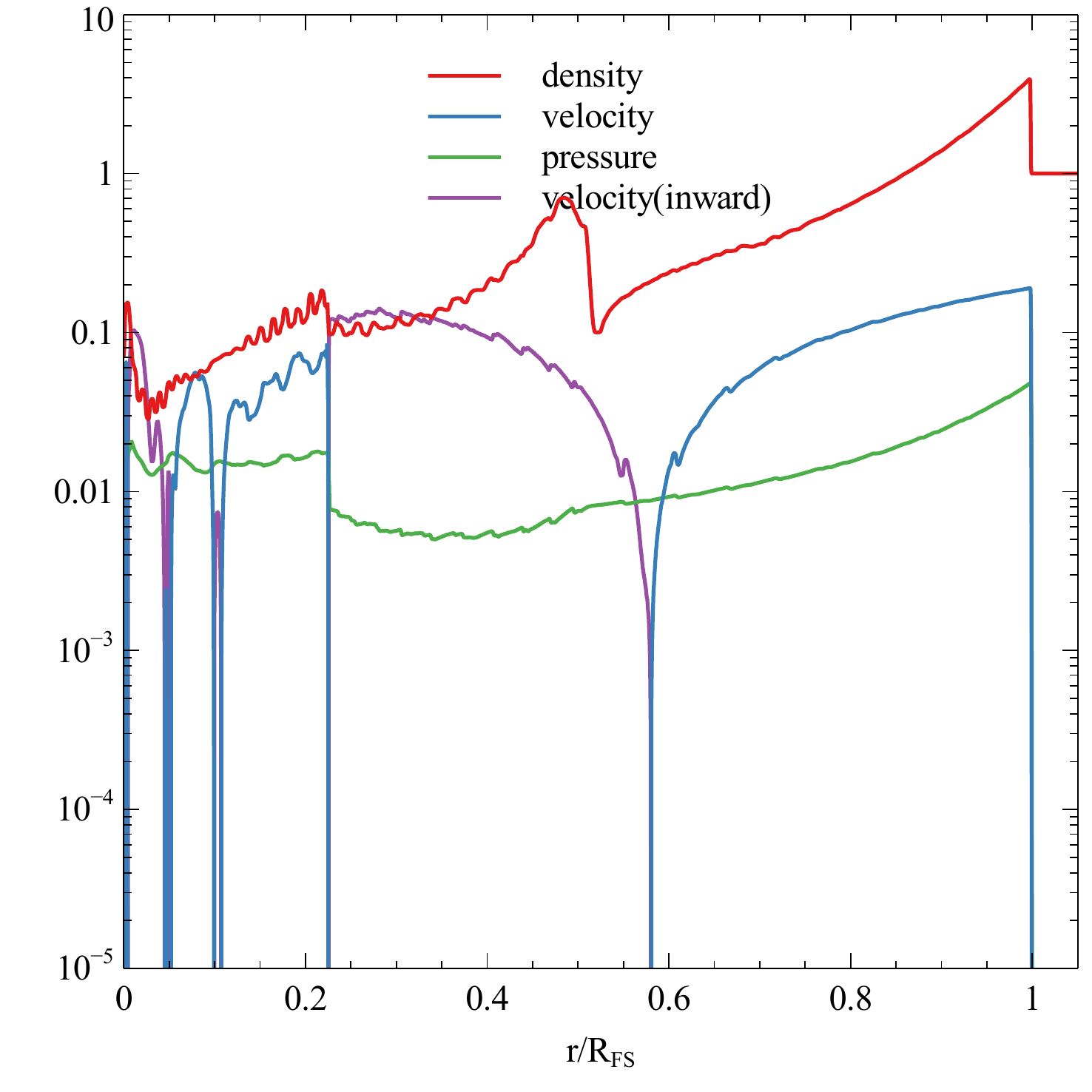}
\includegraphics[angle=0,scale=0.55] {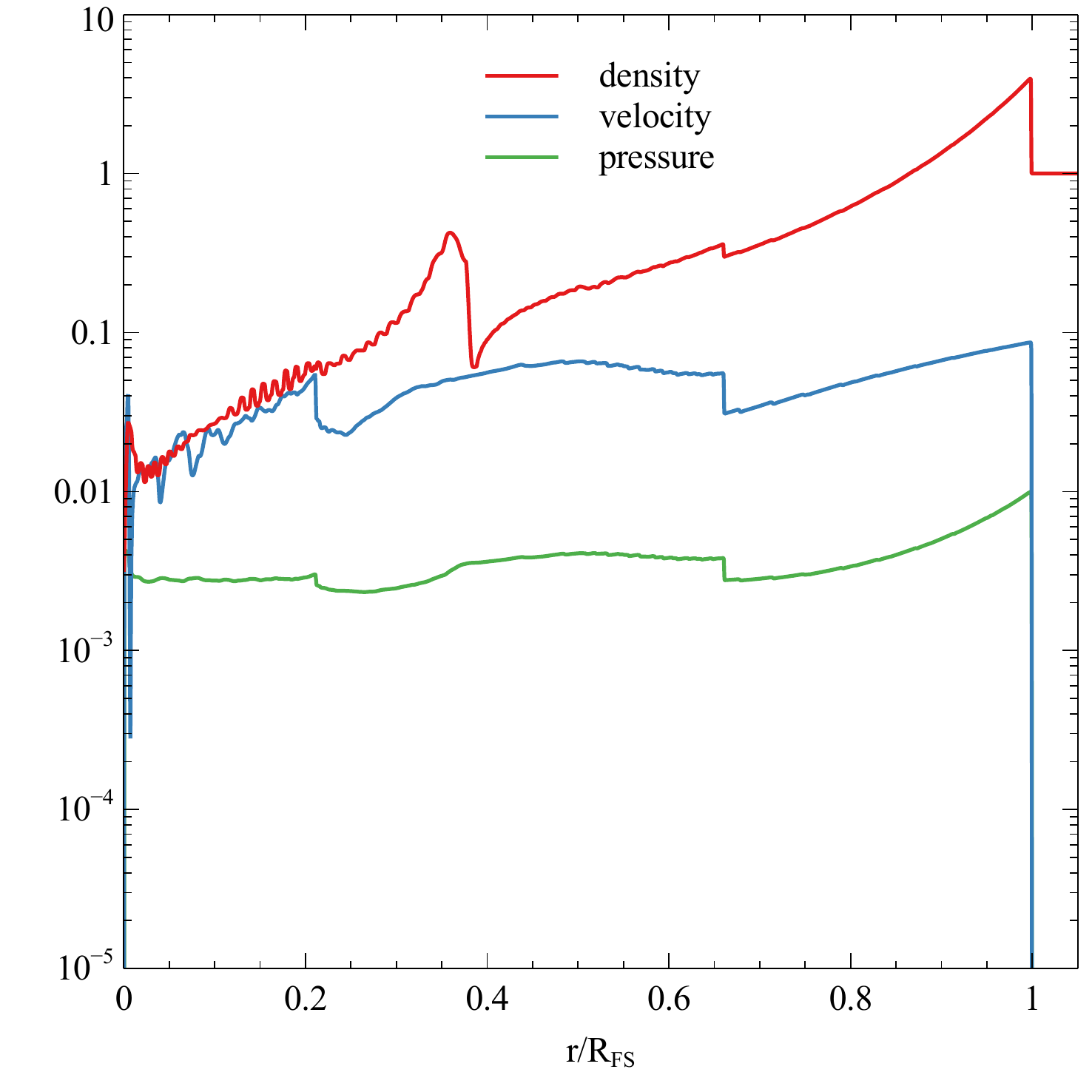}
\caption{
Snapshots of the interior structure for s=0, n=8 from hydrodynamic simulations at 4 characteristic times: $t/t_{ch}\simeq0.10$ 
with $R_{FS}/R_{ch}\simeq0.27$ (top left), 
$t/t_{ch}\simeq1.1$ with $R_{FS}/R_{ch}\simeq1.1$ (top right),  $t/t_{ch}\simeq3.1$ with $R_{FS}/R_{ch}\simeq1.8$ (bottom left) and  $t/t_{ch}\simeq10$ with $R_{FS}/R_{ch}\simeq2.9$ (bottom right). 
%The time  $t/t_{ch}$ and the forward shock radius in units of characteristic  radius ($R_{FS}/R_{ch}$) are given at the top of each panel.
%The density ($rho=\rho/\rho_{ch}$), velocity ($vx=v/v_{ch}$) and pressure ($prs=p/p_{ch}$)
The density, velocity and pressure are scaled to their characteristic values, and are plotted
vs. radius in units of the forward shock radius ($r/R_{FS}$). 
Inward gas velocities are plotted in purple in the bottom left panel.
 \label{fighydrosnaps}}
\end{figure}

The evolution of the interior density, velocity and pressure is captured in the snapshots from PLUTO vs. time. 
Because of the homologous velocity profile of unshocked ejecta ($v\propto r$), the density interior to the RS drops
as $1/t^3$ and the core-envelope boundary expands linearly with time. 
These properties are reproduced by the hydro simulations, with both cases of initial conditions. 
Numerical errors are visible in the results, e.g. the top two panels of Fig.~\ref{fighydrosnaps}. 
The relative errors are largest in unshocked ejecta pressure because the initial pressure is small in that region.  
Errors in density are small except at the origin, and velocity errors are small everywhere.
Case 1 hydro solutions have larger errors than case 2.

Evidence that the solutions are reliable comes from comparison of the solutions with different initial conditions.
Despite large differences in the initial conditions, both case 1 and case 2 evolve to the same structure after time of $t\simeq0.01 t_{ch}$.
The $t<0.01 t_{ch}$ differences between the  case 1 hydro solution and the case 2 hydro solution 
 can be attributed to the inaccuracy of the case 1 initial conditions. 
%For example, the self-similar case 2 n=8 solution has $R_{FS}/R_{RS}=1.215$, different from $R_{FS}/R_{RS}=13/12$ for case 1 
%initial conditions.
Because the simulations for case 1 and case 2 agree after $t\simeq0.01 t_{ch}$, and the
fluctuations for case 2 are smaller than for case 1, hereafter we use the results from case 2.

The self similar evolution of the interior structure from early times ($t<<t_{ch}$) was verified.
Deviations from self-similar evolution as time increases are expected.
These deviations are apparent starting at $t\simeq 0.3t_{ch}$ when reverse shock propagates inward to reach the
the core boundary.  
After this time $R_{RS}/R_{FS}$ decreases.
This can be seen in the animations of the structure files provided with this paper.
At $t/t_{ch}=1$, $R_{RS}$ is well inside the core (top right panel of Fig.~\ref{fighydrosnaps}), and
a sawtooth shape density forms just inside the CD. 
This sawtooth density at the CD persists for the remainder of the SNR evolution. 

After $t/t_{ch}=1$, the reverse shock accelerates inward, reaching the SNR center at $t/t_{ch}\simeq2.5$,
in agreement (within $\sim10\%$) with the results of TM99. 
%The results of TM99 were based on simpler hydro calculations than those presented currently, so the current results  should be more accurate.
After the RS hits the center, a reflected shock slowly propagates outward. 
In the bottom left panel of Fig.~\ref{fighydrosnaps}, for $t/t_{ch}\simeq3$, the reflected shock is propagating outward and
is visible as the pressure, velocity and density jumps at $r/R_{FS}=0.2$.
The reflected shock reaches $r/R_{FS}=0.7$ at $t/t_{ch}\simeq10$  (bottom right panel of Fig.~\ref{fighydrosnaps}), 
and finally reaches the forward shock at  $t/t_{ch}\simeq80$.
 
\begin{figure}
\plotone{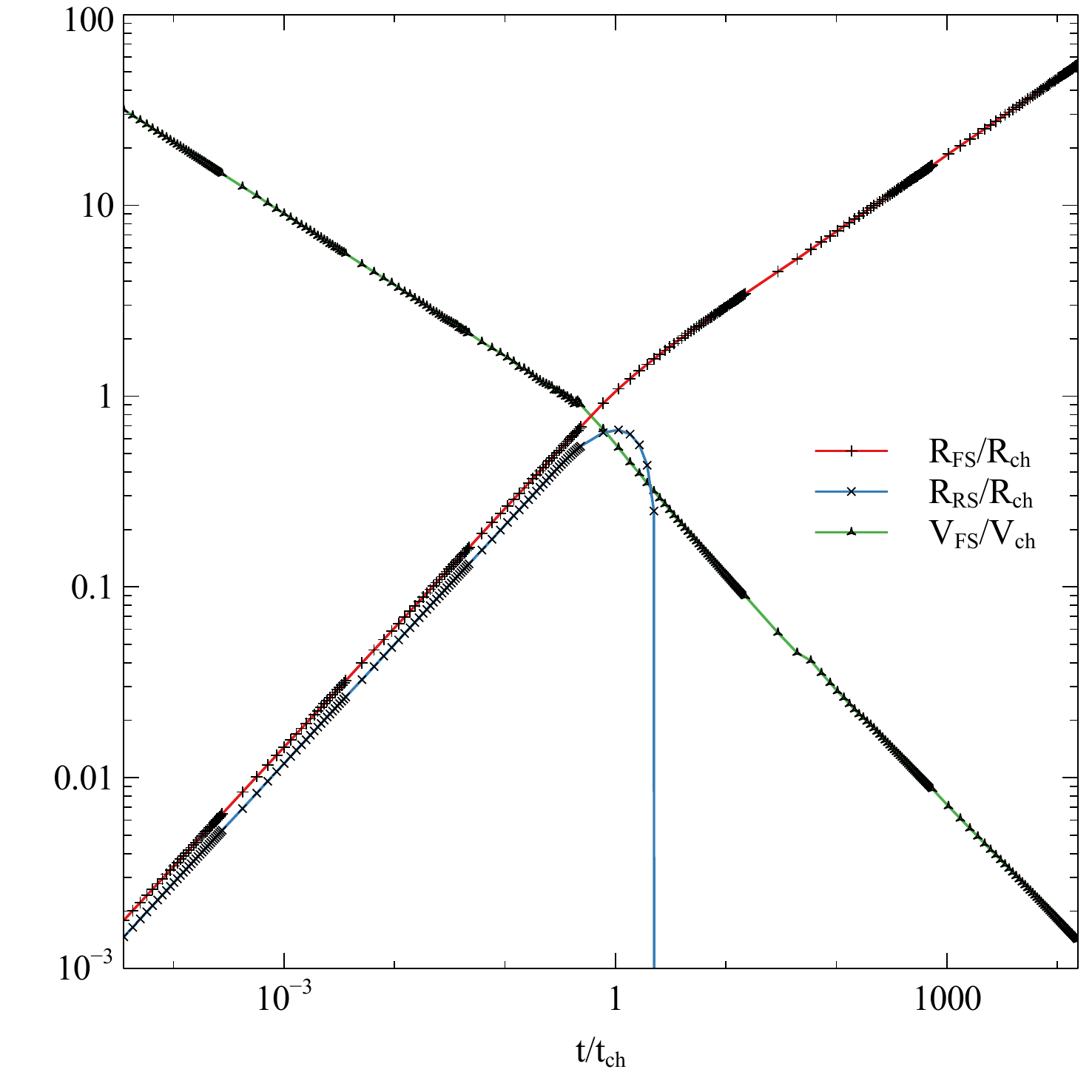}
\caption{
Forward shock radius $R_{FS}$ and velocity $V_{FS}$, and reverse shock radius $R_{RS}$
extracted from the hydrodynamic simulations for s=0, n=8.
Quantities are plotted as in units of characteristic radius or velocity as a function of characteristic time, $t/t_{ch}$. 
 \label{fighydroRV}}
\end{figure}

The evolution for s=0, n=8 of $R_{FS}$, $V_{FS}$ and $R_{RS}$ is shown in Fig.~\ref{fighydroRV}. 
The deviation from self-similar behaviour is seen at $t/t_{ch}\simeq 0.3$.
The reverse shock moves inward after $t/t_{ch}\simeq 1$ and hits the SNR center at $t/t_{ch}\simeq2.5$.
A perturbation in $V_{FS}$ is seen when the reflected shock hits the FS at $t/t_{ch}\simeq80$.
 
Fig.~\ref{fighydroEMT} (left) shows $dEM$ and $dT$ for FS shocked gas for  n=8. 
$dEM_{FS}$ and $dT_{FS}$  vary weakly with time, with maximum range of $\simeq0.8 -0.5$ for $dEM_{FS}$,
and $\simeq1.1 -1.3$ for $dT_{FS}$. 
By comparing different runs with different n, case 1 and case 2 initial conditions, and with other varied parameters, 
and by examining the physical variables in the animations, we determined that the decrease near $t/t_{ch}\simeq 0.6$
and the changes at $t/t_{ch}\simeq 60$ are real. 

In the animations of hydrodynamic variables, one can see the reverse shock encounter the ejecta core at $t/t_{ch}\simeq 0.25$.
Then between $t/t_{ch}\simeq 0.4$ and $\simeq 0.8$, the density profile of the shocked ejecta changes from a 
nearly square-wave shape to a saw-tooth shape  (compare the first two panels in Fig.~\ref{fighydrosnaps}). 
Over the same time interval the density profile of shocked ISM changes from nearly flat just inside the forward shock 
to steeply decreasing inward from the forward shock. 
While this is happening, the pressure profile of both reverse and forward shocked gas changes
from being nearly flat to steeply increasing with radius (compare the first two panels in Fig.~\ref{fighydrosnaps}).
The change in temperature and density profiles combined between $t/t_{ch}\simeq 0.4$ and $\simeq 0.8$
yields an EM-weighted temperature drop compared to the temperature at the forward shock.

At $t/t_{ch}\simeq 2.5$ the reverse shock converges to the center and bounces back outward.
 This reflected shock passes through the CD at $t/t_{ch}\simeq 5$ and reaches the forward shock at $t/t_{ch}\simeq 60$.
It then reflects back inward slowly, not reaching the CD even at the end of the simulations at  $t/t_{ch}\simeq 10^4$.
Meanwhile a second reflected shock, produced from the ejecta density peak at $t/t_{ch}\simeq 5$,  propagates inward
reaching the center at  $t/t_{ch}\simeq 7$. This second reflected shock propagates out to reach the CD at $t/t_{ch}\simeq 13$.
The interior temperature and density profiles of forward-shocked ISM (outside the CD) gradually steepen with
time between $t/t_{ch}\simeq 5$ and  $t/t_{ch}\simeq 60$, but don't change significantly after that.
A drop near $t/t_{ch}\simeq 60$ in $dT_{FS}$ is seen in the first panel of Fig.~\ref{fighydroEMT}.
This is seen in our simulations for all values of n.
This is likely caused by the change in density and pressure profiles caused by the reflected shock encountering the forward shock.
Because the EM-weighted temperature is dominated by the densest gas in a small region around the forward shock, 
it can be significantly affected at the time the reflected shock encounters the forward shock. 

 Fig.~\ref{fighydroEMT} (right) shows $dEM$ and $dT$ for RS shocked gas for  n=8. For RS shocked gas,
$dEM_{RS}$ and $dT_{RS}$ change rapidly with time after the early self-similar phase. 
$dEM_{RS}$ and $dT_{RS}$ both show a significant increase around $t/t_{ch}\simeq 5$, which is real and 
caused by the first reflected shock encountering the density peak in the shocked ejecta.

\subsection{Fits to $dEM$ and $dT$ from the hydrodynamic SNR solutions} 

In order to facilitate usage of $dEM$ and $dT$ for modelling SNRs, we provide fitting functions to 
$dEM$ and $dT$ for both FS and RS gas.
$dEM_{FS}$, $dT_{FS}$, $dEM_{RS}$  and $dT_{RS}$ were extracted from the hydro simulations for n= 6 to 14, 
as functions of scaled time $t_s=t/t_{ch}$. 
We found that piecewise powerlaw functions provide good approximations (i.e. fractional errors less than 5\%) to these quantities.
The minimum number of segments was chosen to give a good fit to the data using least squares minimization.

\begin{figure}
\includegraphics[angle=0,scale=0.55] {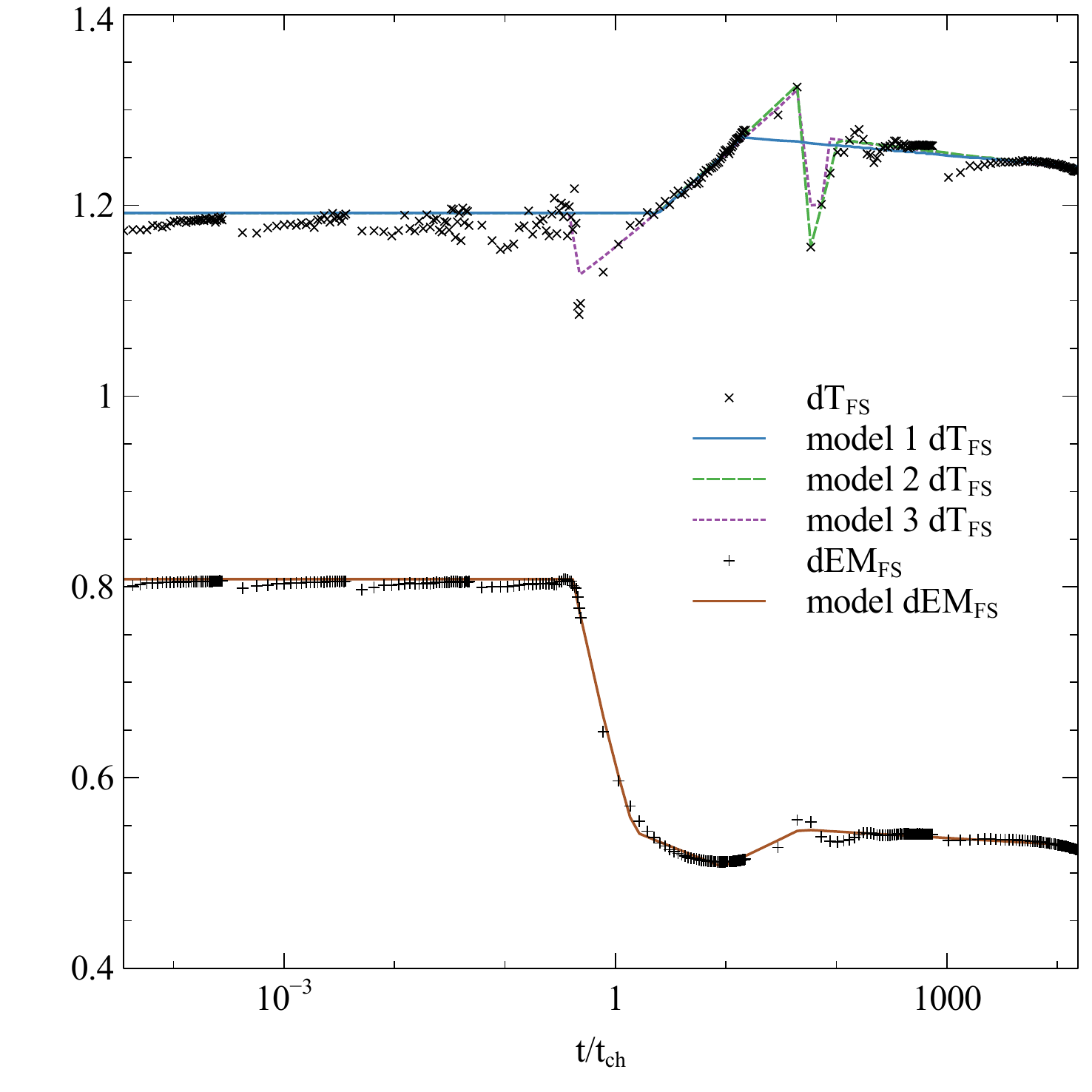}
\includegraphics[angle=0,scale=0.55] {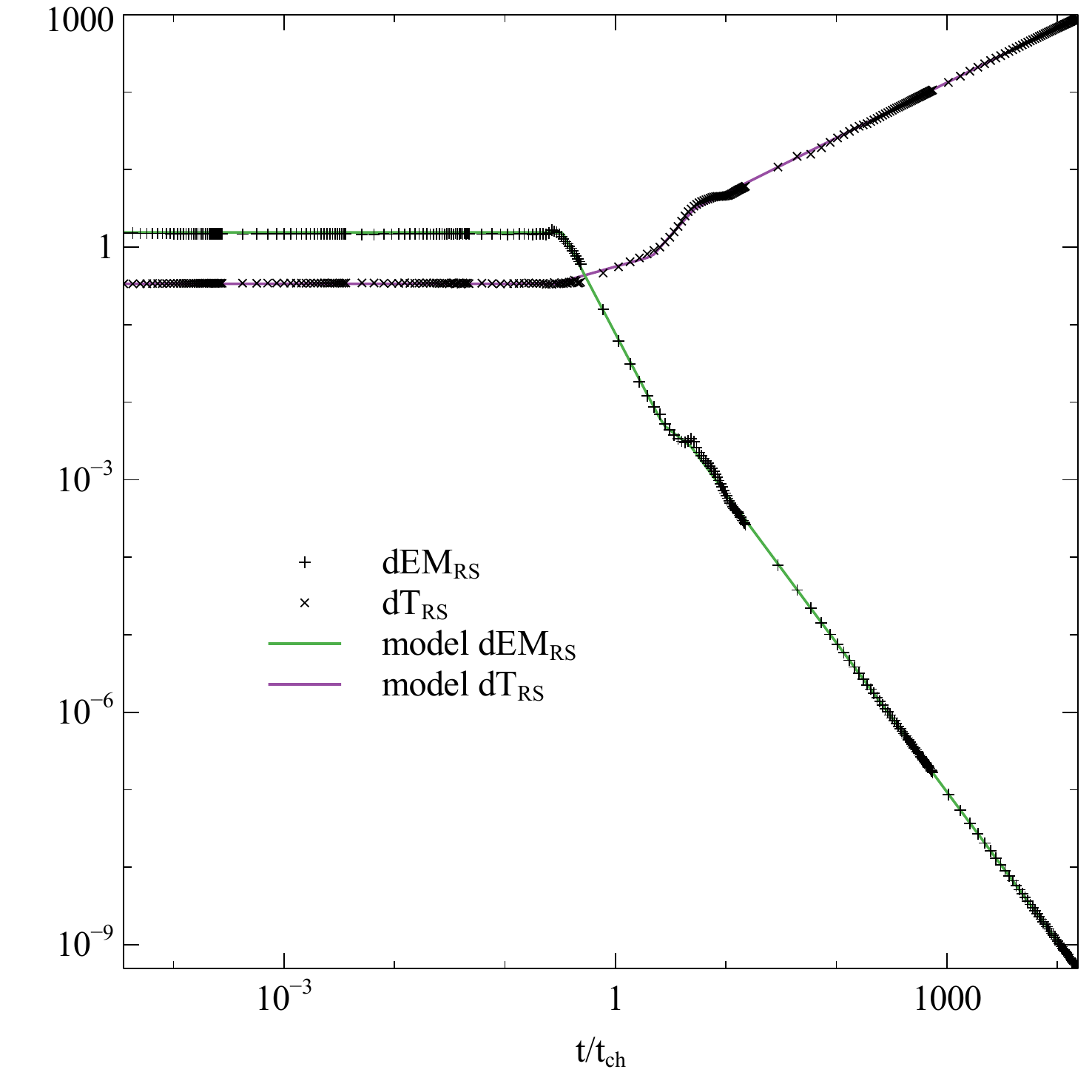}
%\plottwo{hy_EMfTf_n8.pdf}{hy_EMrTr_n8.pdf}
\caption{
Extracted quantities from the hydrodynamic simulations for s=0, n=8 as a function of characteristic time, $t/t_{ch}$.
Left: dimensionless temperature $dT_{FS}$ and dimensionless emission measure $dEM_{FS}$ of forward-shocked gas.  
Right: dimensionless temperature $dT_{RS}$ and dimensionless emission measure $dEM_{RS}$ of reverse-shocked gas. 
The functions fit to  $dT_{FS}$, $dEM_{FS}$, $dT_{RS}$ and $dEM_{RS}$ vs. $t/t_{ch}$ are shown by the lines labelled model. 
 \label{fighydroEMT}}
\end{figure}

Fig.~\ref{fighydroEMT} shows the extracted $dEM$ and $dT$ for n=8 and the n=8 fitting functions.
Because $dT_{FS}$ has the most complex behaviour, we show three different fits: 
one with 6 segments; a second with 5 segments; and a third with 3 segments.
The 6 segment function gives the best fit to the simulation $dT_{FS}$ values, 
and the 3 segment function fits a smoothed version of $dT_{FS}$.
Because real SNRs are not completely spherically symmetric, the RS from different directions 
is expected to encouter the ejecta core and to hit the SNR center at slightly different times. 
Thus a real SNR is expected to have a smoother peak in $dT_{FS}$ at both $t/t_{ch}\simeq0.6$
and $t/t_{ch}\simeq 60$, compared to our hydro simulations. 

We thus chose to use the 3 segment fit to the smoothed $dT_{FS}$ here, for all n values, given by:
\begin{eqnarray}
dT_{FS}(t_s) & =  & dT_{FS,0} \qquad \qquad  \qquad \qquad \qquad ~~~~~  if~~ t_s<t_1 \nonumber \\
&   &   dT_{FS,0} (t_s/t_1)^{p_1} \qquad \qquad \qquad \qquad ~~ if~~ t_1<t_s<t_2 \nonumber \\
&   &   dT_{FS,0} (t_2/t_1)^{p_1} (t_s/t_2)^{p_2} \qquad \qquad ~~~  if~~ t_2<t_s  \nonumber \\
\end{eqnarray}
%\begin{eqnarray}
%dT_{FS} =  & dT_{FS,0} &  if~~ t<t_1 \nonumber \\
%&     dT_{FS,0} (t/t_1)^{p_1} &  if~~ t_1<t<t_2 \nonumber \\
%&     dT_{FS,0} (t_2/t_1)^{p_1} (t/t_2)^{p_2} &  if~~ t_2<t  \nonumber \\
%\end{eqnarray}
For $dEM_{FS}$, a model with 5 segments fits the simulation results:
\begin{eqnarray}
dEM_{FS}(t_s) & =  & dEM_{FS,0} ~~~~~~~~~~~~~~~~~~~~~~~~~~~~~~~~~~~~~~~~~~~~~  if~~t_s<t_1 \nonumber \\
&   &   dEM_{FS,0} (t_s/t_1)^{p_1}~~~~~~~~~~~~~~~~~~~~~~~~~~~~~~~~~~~~ if~~ t_1<t_s<t_2 \nonumber \\
&   &   dEM_{FS,0} (t_2/t_1)^{p_1} (t_s/t_2)^{p_2}~~~~~~~~~~~~~~~~~~~~~~~~~ if~~ t_2<t_s<t_3 \nonumber \\
&    &  dEM_{FS,0} (t_2/t_1)^{p_1} (t_3/t_2)^{p_2} (t_s/t_3)^{p_3} ~~~~~~~~~~~~~~  if~~ t_3<t_s<t_4   \nonumber \\
&    &  dEM_{FS,0} (t_2/t_1)^{p_1}  (t_3/t_2)^{p_2} (t_4/t_3)^{p_3} (t_s/t_4)^{p_4}~~~  if~~ t_4<t_s   \nonumber \\
\end{eqnarray}

 The right panel of  Fig.~\ref{fighydroEMT}  shows $dEM$ and $dT$ for RS shocked gas.
 After  $t/t_{ch}\sim0.3$, $dEM_{RS}$ decreases and $dT_{RS}$ increases. 
 The main cause of the decrease of $dEM_{RS}$ is the increase of volume of FS gas (via $R_{FS}$) relative to volume of RS gas (see equation 2).
The main cause of the increase of $dT_{RS}$ is the decrease of $T_{FS}$ relative to slower decrease of $T$ of RS gas (see equation 4).

For $dT_{RS}$ and for $dEM_{RS}$, 4 segments fit the simulation results:
\begin{eqnarray}
dT_{RS}(t_s)&=&  dT_{RS,0} ~~~~~~~~~~~~~~~~~~~~~~~~~~~~~~~~~~~~~~~~~ if~~t_s<t_1 \nonumber \\
&   &   dT_{RS,0} (t_s/t_1)^{p_1} ~~~~~~~~~~~~~~~~~~~~~~~~~~~~~~~ if~~ t_1<t_s<t_2 \nonumber \\
&   &  dT_{RS,0} (t_2/t_1)^{p_1} (t_s/t_2)^{p_2} ~~~~~~~~~~~~~~~~~~~~~  if~~ t_2<t_s<t_3   \nonumber \\
&   &  dT_{RS,0} (t_2/t_1)^{p_1} (t_3/t_2)^{p_2} (t_s/t_3)^{p_3} ~~~~~~~~~~~  if~~ t_3<t_s   \nonumber \\
\end{eqnarray}.
\begin{eqnarray}
dEM_{RS}(t_s)&=&  dEM_{RS,0} ~~~~~~~~~~~~~~~~~~~~~~~~~~~~~~~~~~~~~~~~ if~~t_s<t_1 \nonumber \\
&   &   dEM_{RS,0} (t_s/t_1)^{p_1} ~~~~~~~~~~~~~~~~~~~~~~~~~~~~~~~ if~~ t_1<t_s<t_2 \nonumber \\
&   &  dEM_{RS,0} (t_2/t_1)^{p_1} (t_s/t_2)^{p_2} ~~~~~~~~~~~~~~~~~~~~~  if~~ t_2<t_s<t_3   \nonumber \\
&   &  dEM_{RS,0} (t_2/t_1)^{p_1} (t_3/t_2)^{p_2} (t_s/t_3)^{p_3} ~~~~~~~~~~  if~~ t_3<t_s  \nonumber \\
\end{eqnarray}

The best fit coefficients for $dT_{FS}(t_s)$, $dEM_{FS}(t_s)$, $dT_{RS}(t_s)$ and $dEM_{RS}(t_s)$  are given in Table~\ref{tab:tabhd}
for the different values of n. 
$dT_{FS,0}$, $dEM_{RS,0}$, $dT_{FS,0}$ and $dEM_{RS,0}$  were fixed at the values for
the initial CP self-similar phase for s=0 and the different n values.

In general terms, the dividing times between the segments of nearly powerlaw behaviour correspond to
changes in the interior physical structure, either pressure or density, of the SNR. 
The first segment covers the self-similar early ED phase, and ends near, or soon after, the time
when the inward moving (in the frame of the ejecta) reverse shock encounters the power-law core of the ejecta.
For the shocked-ISM quantity $dEM_{FS}$, the end of the second segment is near the time that the
density and pressure profiles of the shocked-ISM steepen (seen in the animations). 
The end of the third segment is near the time that the reflected shock propagates to the inner boundary of the shocked-ISM
and starts to flatten the density and pressure profiles.
The end of the fourth segment is when the reflected shock reaches the forward shock at $t/t_{ch}\simeq 50$.
The fifth segment lasts from that time onward.

The shocked-ISM quantity, $dT_{FS}$, the simplified 3 segment fits shown here have two transition times. 
The physical transitions times are as noted in the above paragraph. 
The best-fit transition times are phenomenological.
The first is intermediate between the time of reverse shock encountering the ejecta core and when it reaches the center.
The second  is intermediate between the time of reflected shock encountering the inner edge of shocked ISM and 
the time when it encouters the forward shock.

For the shocked-ejecta quantities, $dEM_{RS}$  and $dT_{RS}$, the end of the second segment is near the time that the
reverse shock reaches the SNR center. 
The end of the third segment is near the time that the reflected shock propagates
outward to reach the peak in density of the shocked ejecta. 
The fourth segment lasts from that time, $t/t_{ch}\simeq 5$, onward.

With the calculated time-dependent $dEM$ and $dT$, we can compare how the properties of the shocked ISM 
and shocked ejecta change with time. 
The FS EM and EM-weighted T, in dimensionless form, remain remarkably constant over the whole SNR evolution:
From  Fig.~\ref{fighydroEMT} (left), we see that $dT_{FS}$ only rises a small amount ($\sim10$\%) between 
$t/t_{ch}$ = 1 and 10, whereas significant changes in $dT_{RS}$ and $dEM_{RS}$ occur. 
The very slow decrease of $dT_{FS}$ after $t/t_{ch}\sim100$ occurs in all the simulations, providing evidence it is a real effect. 
This decrease is probably caused by the effect of the reflected shock on expanding the forward shock 
more than it does in the self-similar solution. 
$dEM_{FS}$ exhibits small changes, with a sharp decrease of $\sim40$\% between $t/t_{ch}$ = 0.4 and 2,
during the time the reverse shock propagate through the ejecta core.

For shocked ejecta, $dT_{RS}$ rises steadily with time after the self-similar phase ends, near
$t/t_{ch}$ = 0.3, and has a bump around $t/t_{ch}\simeq$6, when the reflected shock passes through the 
dense shell of ejecta near the CD.  
We find that the shocked ejecta is hotter than the shocked ISM ($dT_{RS}> dT_{FS}$)
for $t/t_{ch}\gtrsim$ 2 for n=6. 
This transition of  $T_{RS}> T_{FS}$
(noting that the scaling is the same for both values to obtain dimensionless values)
gradually increases from $t/t_{ch}\gtrsim$ 2 for n=6 to $t/t_{ch}\gtrsim$ 4 for n=14.

$dEM_{RS}$ (right panel of Fig.~\ref{fighydroEMT}) decreases steadily with time after the self-similar phase ends, near
$t/t_{ch}$ = 0.3, and has a bump around $t/t_{ch}\simeq$ 6, caused by the reflected shock.
The $EM$ of shocked ejecta is smaller than that of shocked ISM  ($dEM_{RS}< dEM_{FS}$) at all times for
n=6 and 7. For n=8 to 14, $dEM_{RS}>dEM_{FS}$ at early times and $dEM_{RS}<< dEM_{FS}$ at late times.
The transition time for the $EM$ of shocked ISM to exceed that of shocked ejecta increases from
 $t/t_{ch}$ = 0.4 for n=8 to  $t/t_{ch}$ = 0.6 for n=14.

There are other small changes in $dT_{FS}$, $dEM_{FS}$ , $dT_{RS}$ and $dEM_{RS}$. 
Comparison with the hydro simulations snapshots shows
that these are caused by the RS shock encountering then propagating rapidly through the ejecta core, 
reflecting from
convergence at the SNR center, and then propagating outward through the shocked ejecta, the CD and 
the shocked ISM to the FS. 
%To illustrate the above changes, an animation of $dT_{FS}(t_s)$, $dEM_{FS}(t_s)$, $dT_{RS}(t_s)$ and $dEM_{RS}(t_s)$
% vs. time with animation parameter n is provided as an online attachment.

\subsection{Application of the Model to SNR Observations} 

In order to demonstrate the current model, we apply it to three historical Galactic SNRs with published electron temperatures and emission measures: Kepler, Tycho and SN1006. 
The model is applied to the FS emission of each SNR to test if the model can reproduce the observed FS radius, FS emission measure, FS temperature
and the known SNR age. 
Table~\ref{tab:snrdata} gives the observed age, type, radius, FS emission measure and FS temperature for these SNRs. 
All three SNRs are Type Ia, so we use an estimated ejected mass of 1.2$\textrm{M}_\sun$, 
and also test models with 1.0$\textrm{M}_\sun$ and 1.4$\textrm{M}_\sun$. 
In the following sections, we discuss the model results and implications. 

Table~\ref{tab:snrdata2} gives the observed shocked ejecta (RS) emission measures and temperatures. 
Kepler and Tycho each have three measured ejecta components and SN1006 has two measured ejecta components. 
The observations indicate that layering of the ejecta is important in all three SNRs. 
The current model is simplified and assumes the ejecta is uniformly mixed. 
%Layering could be implemented in the current model but results in significant complication of the model. 
We do not explore the complication of ejecta layering in the current model but leave that to future work. 

\subsubsection{G$4.5+6.8$ (Kepler)}

The remnant of the Kepler's SN of 1604 is a well studied SNR. 
\cite{1999Kinugasa} analyse the relative abundances of the ejecta to classify Kepler's SN as a Type Ia SN.
The distance to the $3^{\prime}$ diameter SNR has been a difficult obtain. 
The estimates range from a lower limit of 3.0 kpc \citep{2005Sankrit} to an upper limit of 6.4 kpc \citep{1999Reynoso}. 
For this study we use the distance of  $ 5.1^{+0.8}_{-0.7}$ kpc presented by \cite{2016Sankrit} using improved proper motion measurements and revised values of shock velocities.

We show in Table~\ref{tab:snrmodels} a set of models that reproduce the measured radius, FS emission measure and FS temperature for Kepler.
The models are listed with the assumed SNR distance and model input parameters (s, n and ejected mass). 
The model outputs are SNR age and explosion energy, ISM density (for s=0) or wind parameter $\rho_s$ (for s=2), RS emission measure, RS
temperature and RS radius.  
The models for a SNR in a uniform ISM give an age between 1360 and 1680 yr, for the full range of n=6 to 14, 
far too large compared to the real age of 407 yr. 
 However, models for a SNR in a stellar wind (s=2) give much lower ages, between 102 and 332 yr, for the full range of n=6 to 14. 
 The s=2 n=6 model gives an age closest to the observed age.
 Varying the ejecta mass for the s=2 n=6 model does not alter the age. 

 The s=2, n=6 model using the upper distance estimate (5.9 kpc) give an age closer to the observed age.
 This d=5.9 kpc model has $EM_{RS}\simeq3.2\times10^{54}$cm$^{-3}$, 
 which agrees %(to 2 digit accuracy) 
 with the sum of the 3 observed $EM_{RS}$ components. % of $\simeq3.2\times10^{54}$cm$^{-3}$. 
 The model $kT_{RS}$ of 3.55 keV is somewhat higher than observed value of 2.59 keV for the dominant ejecta component. 
This could be caused by the model's assumption that the RS heated gas has
 the same $T_e$ to $T_{ion}$ ratio\footnote{ $T_e/T_{ion}$ is calculated using Coulomb heating, see LW17 for details.} 
 as the FS heated gas.
 The model explosion energy is $0.89\times10^{51}$ erg. 
 The stellar wind parameter gives the mass loss rate divided by wind velocity, so that for a wind velocity of 10 km/s, the inferred
 mass-loss rate is  $\simeq7\times10^{-6}$M$_{\odot}$/yr.
 This is consistent with expected values for a red giant star, as a companion to an accreting white dwarf as the progenitor system.
 
 Kepler's SNR was studied using hydrodynamic simulations combined with X-ray spectral synthesis by \citet{2012Patnaude} .
 Two ejecta models were used as input: DDTa, with explosion energy $1.4\times10^{51}$erg and DDTg with explosion energy $0.9\times10^{51}$erg \citep{2003Badenes}.
The forward shock evolution and the associated X-ray spectra for different CSM environments were calculated and compared to 
the Chandra spectrum of a southern pie-shaped region of the SNR. 
They required the simulated X-ray spectrum to produce the observed Si, S and Fe line centroid energies and line ratios.
For an SNR in a wind CSM (s=2) the DDTa and DDTg models were ruled out. 
The DDTa models were consistent if they included a central cavity with radius $\sim10^{17}$ cm inside the wind, resulting 
in distance $>$7 kpc. 
For an SNR in a constant density CSM (s=0) the DDTa models gave spectra consistent with observations for distance 5-6.5 kpc.
But the s=0 models were noted to be inconsistent with the X-ray morphology of Kepler's SNR.

Now we compare our results to those of \citet{2012Patnaude}. 
Instead of 2 specific ejecta models, our models allow a continuous range of energies and explosion masses, 
and variable ejecta density profiles (n=6 to 14). 
Instead of fitting line centroids and line ratios, we fit the total EM of the forward shock. 
We take distance as input and require model age to be close to the observed age, instead of taking age as  input and 
distance as output.
Neither approach considers non-sphericity of the ejecta nor of the CSM.
The two approaches are very different, yet the conclusions are similar: s=0 models require small distances, and s=2 models require
larger distances. 
Our model shows, using the new $EM_{FS}$ from \citet{2015Katsuda}, the s=0 models are ruled out. 
We could not test our s=2 n=6 d=5.9 kpc model for line centroids and line ratios, but the results of \citet{2012Patnaude} on the 
line centroids and ratios likely imply that that the s=2 stellar wind has a central cavity.

\subsubsection{G$120.1 +1.4$ (Tycho)}

Tycho (G$120.1 +1.4$) is a  SNR %with a $4^{\prime}$ radius and 
first studied by Tycho Brahe in 1572. 
Examining the light-echo spectrum, \cite{2008Krause} provided evidence for it to be a Type Ia.  
The distance has been determined to be 1.7 \citep{1986Albinson} and 3 to 5 kpc \citep{2010Hayato}. 
We use the latter distance of $4.0 \pm 1.0$ kpc. 

In Table~\ref{tab:snrmodels} models that reproduce the measured radius, FS emission measure and FS temperature for Tycho are given.
Models for a SNR in a uniform ISM give an age far too large  ($\sim$4000 yr) compared to the real age of 434 yr. 
This supports models for a SNR in a stellar wind (s=2). The s=2 n=7 model gives an age of 378 yr roughly consistent with the observed age.
%Varying the ejecta mass for the s=2 n=6 model does not alter the age. 
 Models using different distance estimates show that better agreement  with the observed age is obtained for 
 the s=2 n=7  distance 4.5 kpc model (age of 428 yr).
An alternate way of obtaining the observed age is to use an s=2 n=8 model and decrease the distance to $\simeq$3 kpc, 
as shown in Table~\ref{tab:snrmodels}.
 The s=2, n=7, d=4.5 kpc model explosion energy is large, $3.86\times10^{51}$ erg, which indicates that
 the s=2, n=8, d=3 kpc model, with explosion energy is $0.85\times10^{51}$ erg, is more realistic.
 Another possiblity is  an ejecta profile intermediate between n=7 and n=8, which is not in the current model, and SNR distance between 3 and 4 kpc.
%That model still has a slightly too large (498 yr)

 The s=2, n=8, d=3 kpc model has $EM_{RS}\simeq2\times10^{55}$cm$^{-3}$, compared to the sum of 
 the 3 observed $EM_{RS}$ components of $\simeq2.2\times10^{54}$cm$^{-3}$ at d=4.5 kpc. 
 This could be caused by incorrect estimates of the ejecta composition in the model, given in the footnote of  Table~\ref{tab:snrmodels}. 
 The model $EM_{RS}$ is sensitive to the ratio of heavy elements to hydrogen in the ejecta, so the model $EM_{RS}$
 could be adjusted to agree with the measured $EM_{RS}$ by decreasing the ejecta hydrogen abundance by a factor of several. 
% The model $kT_{RS}$ is higher than observed, which could be caused by the model's assumption that the RS heated gas has
% the same electron-temperature to ion-temperature ratio as the FS heated gas.
 The stellar wind parameter for either  s=2, n=7, d=4.5 kpc or  s=2, n=8, d=3 kpc models 
 gives, for a wind velocity of 10 km/s, an inferred mass-loss rate of $\sim6\times10^{-6}$M$_{\odot}$/yr,
  consistent with expected values for a red giant star.
 
Tycho's SNR was studied using hydrodynamic simulations combined with X-ray spectral synthesis by \citet{2006Badenes} .
Several Type Ia explosion models were used as input.
The X-ray spectra were calculated for different ISM densities ($\rho_{amb}$) and different electron-to-ion internal energies 
($\beta=\epsilon_{e}/\epsilon_{ion}$) and compared to the XMM-Newton spectrum of an eastern pie-shaped region of the SNR. 
They required the model X-ray spectrum to produce the observed Si, S and Fe line centroid energies and line ratios of
the emission from the ejecta. 
Only delayed detonation models were found to give consistency with the observed lines.
The DDTc model was best, with $\rho_{amb}=2\times10^{-24}$g/cm$^{3}$ and $\beta=0.03$.
DDTc had explosion energy of $0.85\times10^{51}$ erg and gave a model distance to Tycho of 2.59 kpc.
  
Another approach to modelling Tycho's SNR is exemplified by \citet{2014Slane}. 
The broadband ($10^{-8}$eV to  $10^{15}$eV) spectrum, the FS radius, and X-ray and radio surface brightness 
profiles are modelled, using hydrodynamic simulations with a semi-analytic treatment for diffusive shock acceleration. 
Their main conclusions were that the ambient medium density is $\sim0.3$ cm$^{-3}$, ambient magnetic field is $\sim0.5~\mu$G,
$\sim$16\% of the kinetic energy is in relavistic particles with electron energy to proton energy ratio of $\sim0.003$,
and the distance is $\sim3.2$ kpc.

We compare our results to the above. % then of \citet{2014Slane}. 
 \citet{2006Badenes} did not model the forward shock emission, nor consider a wind-type CSM.
Instead of specific ejecta models and fixed $\rho_{amb}$ and $\beta$ values, 
our models have a continuous range of energies and explosion masses, variable ejecta density profiles (n=6 to 14), variable
ambient density and $\beta$ determined by the Coulomb heating model. 
Instead of fitting line centroids and line ratios, we fit the EM, temperature and radius of the forward shock. 
We take distance as input and require model age to be close to the observed age.
Our model, using the new $EM_{FS}$ from \citet{2015Katsuda}, rules out the s=0 models. 
This conclusion assumes the evolution of the FS is not strongly altered by energy losses to cosmic ray acceleration. 
However, the results of \citet{2014Slane} indicate that the energy in cosmic rays is significant.
The three different approaches have very different assumptions, and none considers non-sphericity of the ejecta nor of the CSM.
%Our model fits the bulk emission  of the FS, whereas  \citet{2006Badenes} fits the emission lines from the RS. 
They are sensitive to different aspects of the SNR and its ejecta, and all probably correctly capture different aspects of the SNR.

\subsubsection{G$327.6 +14.6$ (SN1006)}

The historic supernova SN1006 is likely the brightest stellar event recorded \citep{1966Minkowski}. 
\cite{1996Schaefer} presented an argument that SN1006 was a Type 1a.  
%The  remnant is $15^{\prime}$ in diameter. 
The proper motion measurements combined with the shock velocity, yield a  distance to the SNR of $2.18 \pm 0.8$ kpc \citep{2003Winkler}.  

Models that reproduce the measured radius, FS emission measure and FS temperature for SN1006 are given in Table~\ref{tab:snrmodels} .
The SNR in a uniform ISM gives an age far too large ($\sim$8000 yr) compared to the real age of 1002 yr. 
This supports models for a SNR in a stellar wind (s=2).
 For d=2.18 kpc, the s=2 n=8 and s=2 n=9 models yield ages just below and just above the observed age.
%Varying the ejecta mass for the s=2 n=6 model does not alter the age. 
 We tested models with the upper and lower distance values. 
The d=2.26 kpc s=2 n=8 model age of 957 yr is closest to the real age
%, whereas the d=2.10 kpc s=2 n=9 model age of 1095 yr is similarly different from the as the  d=2.18 kpc, the s=2 n=8 age.
% match the observed age, nor does the lower distance limit decrease the n=9 model age enough.
 Varying the ejecta mass does not affect the model age.
We adopt the d=2.26 kpc, the s=2 n=8 model as the best match to SN1006.
 
 The  model $EM_{RS}\simeq5\times10^{53}$cm$^{-3}$ is larger than the sum of 
 the 2 observed $EM_{RS}$ components of $\simeq9\times10^{52}$cm$^{-3}$.
As noted for Tycho's SNR above, the model ejecta composition could be adjusted to obtain agreement.
 The model $kT_{RS}=2.1$ keV is higher than the observed $kT_{RS}$ and is dependent on the ejecta composition and electron to ion temperature ratio, which are not explored in detail here.
% The model $kT_{RS}$ is higher than observed, which could be caused by the model's assumption that the RS heated gas has
% the same electron-temperature to ion-temperature ratio as the FS heated gas.
 The inferred explosion energy is $1.36\times10^{51}$ erg, consistent with a bright explosion.
 The stellar wind parameter is  $1.6\times10^{-7}$M$_{\odot}$s/(km yr). 
 Thus a wind velocity of 10 km/s, yields an inferred mass-loss rate of $\simeq2\times10^{-6}$M$_{\odot}$/yr,
  consistent with values for a red giant star.
  
Recent models for SN1006 are given by \citet{2018MR}. 
They use a set of 8 Type Ia explosion models in a constant density ISM and with $T_e/T_{ion}$ fixed at the low value of $m_e/m_{ion}$.
A hydrodynamics code, without diffusive shock acceleration, is used to compute SNR evolution and the resulting
centroid energy and luminosity of the Fe K$\alpha$ line. For SN1006 (their Fig. 14) they find that the observed radius is much 
larger than any of the model radii. 
This supports our conclusion that an SNR with s=0 cannot explain SN1006, and that a stellar wind (s=2) CSM is required.

 In summary, we have compared the results of our simplified models to more detailed models for three historical SNRs.
 Our model was designed to fit the bulk properties of a SNR: its EM, temperature and radius, whereas different detailed models
 were designed to fit other aspects of a SNR. 
 These include: modelling of one or several of the emission lines from the ejecta, which gives valuable information about the
 ejecta properties; and modelling the broadband (radio, X-ray and gamma-ray) spectrum of an SNR, which gives valuable
 information about the shock acceleration process.
 Another point of consideration is the ejecta density profile. We have used a power-law density profile because that 
 is consistent with the unified evolution of TM99, and associated scaling of the models.
The ejecta profiles associated with specific explosion models, as discussed above in relation to detailed models for Kepler,
Tycho and SN1006 do not give rise to unified evolution. 
Another possibility is discussed by \citet{1998Dwarkakas} who argue that SN  Ia ejecta are better approximated
by exponential profiles than power-law profiles. We will explore this in future work.
%We do not explore this because it also does not give rise to unified evolution.
 As noted earlier, the purpose of our simplified model is to quickly give estimates of the SNR explosion energy, age, 
 ejecta mass and density profile, and ISM density or CSM mass-loss parameter. 
 These values are intended to be used as approximate inputs to guide more detailed modelling for specific SNRs where data allows 
 further modelling of line energies and strengths or the broadband spectrum. 
 Most SNRs have only limited data, and for those SNRs, the simplified model we have presented here is still applicable. 

\section{Summary}

Quantitative calculations of $EM$s and $EM$-weighted temperatures are needed in order to model X-ray emission from SNRs.
A comprehensive and easy-to-use set of $EM$s and $EM$-weighted temperatures have not been presented before.
We present such results for the case of spherically symmetric SNR evolution.
Such SNRs undergo a unified evolution, as demonstrated by TM99, from explosion to the onset of significant radiative losses.
The unified evolution means that the SNR structure, expressed in dimensionless variables, is a function of dimensionless time only.
The dependence on explosion energy, ejecta mass and ISM density is built into the definitions of the characteristic variables.
Thus a single set of time-evolved models suffices to calculate a general set of $EM$s and $EM$-weighted temperatures. 

The early ED evolution is self-similar. The associated CP solutions were recalculated.  
%Tables of the solutions are provided as online attachments, for s=0 and s=2 and all n from 6 to 14, 
%with resolution of 500 points between RS and FS.
The dimensionless column EM, $cEM$, for these solutions are presented in  Fig.~\ref{figs0cem}  and Fig.~\ref{figs2cem},
and included in the CP solution tables.
The summary quantities $dT_{FS}$, $dEM_{FS}$, $dT_{RS}$ and $dEM_{RS}$ are given in Table~\ref{tab:tabss}.
For non-radiative SNRs with the ejecta mass much smaller than the swept-up ISM mass, the evolution is self-similar.
The WL91 solutions are  recalculated here for the cases of uniform ISM ($C/\tau$=0) and cloudy ISM ($C/\tau$=1, 2 and 4).
The $C/\tau$=0 case is the same as the pure Sedov-Taylor solution.
%Tables of the solutions and $cEM$s are  provided as online attachments.
The summary quantities $dT_{FS}$ and $dEM_{FS}$ are given in Table~\ref{tab:tabss}.
 
We calculated the unified phase evolution using the publicly available hydrodynamic code PLUTO \citep{2007mignone}
for s=0 and n values from 6 to 14. The final results presented here use the CP solutions as initial conditions.
The evolution is calculated from early times ($t/t_{ch}\sim10^{-4}$) to late times ($t/t_{ch}\sim10^{4}$), 
while maintaining a minimum resolution of the SNR of 1000 grid points center to FS.
We verified that the early-time hydro solutions ($t/t_{ch}\lesssim0.2$) exhibit self-similar evolution agreeing with the CP solutions. 
%Animations of the SNR with $\sim250$ timesteps each are made available as online attachments. 
These illustrate the changes in interior structure as the SNR evolves. 

Summary quantities $dT_{FS}$, $dEM_{FS}$, $dT_{RS}$ and $dEM_{RS}$ were calculated as a function of time. 
Those for n=8 are presented in Fig.~\ref{fighydroEMT} as an example.
% and an online attachment shows these functions for all values of n. 
The other values of n show similar behaviour to
that seen for n=8, including the significant changes caused by the RS accelerating toward the center, the
RS reflecting off of the center, and the RS passing through the material concentrated near the CD and the FS.
The latter change occurs at the late time of $t/t_{ch}\sim60$.
%Model emission measures and temperatures, required fitting X-ray observations of SNRs, were calculated from the simulations.\
Piecewise powerlaws were least-squares fit to the dimensionless emission measures and temperatures, 
$dT_{FS}(t/t_{ch})$, $dEM_{FS}(t/t_{ch})$, $dT_{RS}(t/t_{ch})$ and $dEM_{RS}(t/t_{ch})$. 
The powerlaws are given by equations (7) to (10) with coefficients given in Table~\ref{tab:tabhd}
for the different values of n. 
The emission measure and temperature fitting functions can be used in modelling SNRs, without the need to run the hydrodynamic simulations. 

Real SNRs in our Galaxy (or in other galaxies) have asymmetric ejecta to various degrees and expand in non-homogeneous media.
$EM$s and $EM$-weighted temperatures from spherically symmetric models can yield approximate values of
SNR age, explosion energy, ejecta mass and ISM density.
In cases where observations are detailed enough to warrant more detailed study, estimates of SNR parameters from the current spherically symmetric evolution can be used as inputs for 2-D or 3-D hydrodyamic simulations.

We illustrate the application of our models to three historical SNRs: Kepler, Tycho and SN1006.
The main results from matching the models to the observed radius, age, $T_{FS}$ and $EM_{FS}$ are as follows. 
All three of these Type Ia SNRs are consistent with expansion in a stellar wind environment but not a constant density ISM 
and the inferred wind parameters are consistent with red giant companions.
The inferred explosion energies for all three are between $0.85\times10^{51}$ and $1.4\times10^{51}$ erg. 

\subsection{Software Release}

\begin{figure}
\includegraphics[angle=0,scale=0.45] {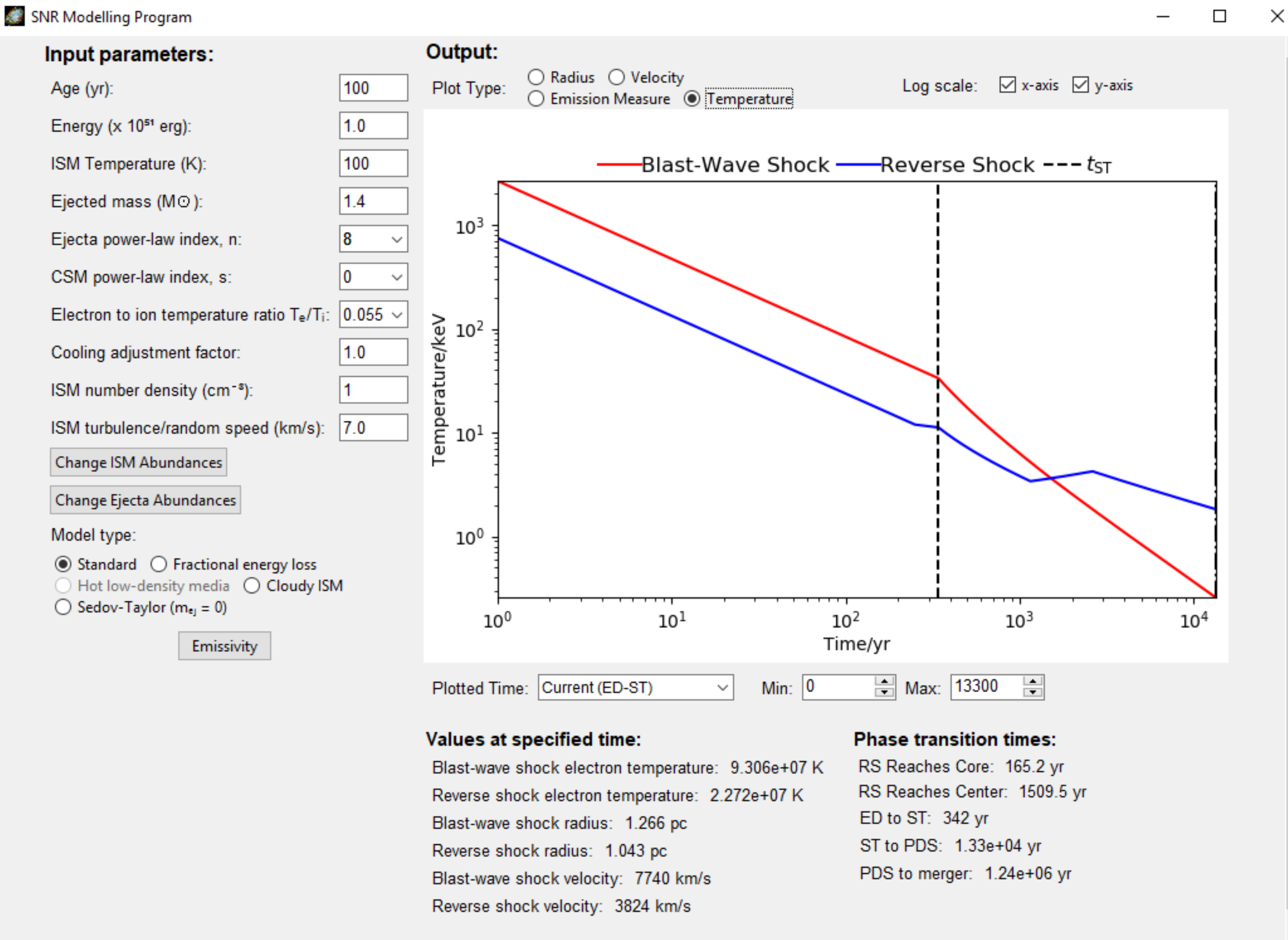}
\includegraphics[angle=0,scale=0.45] {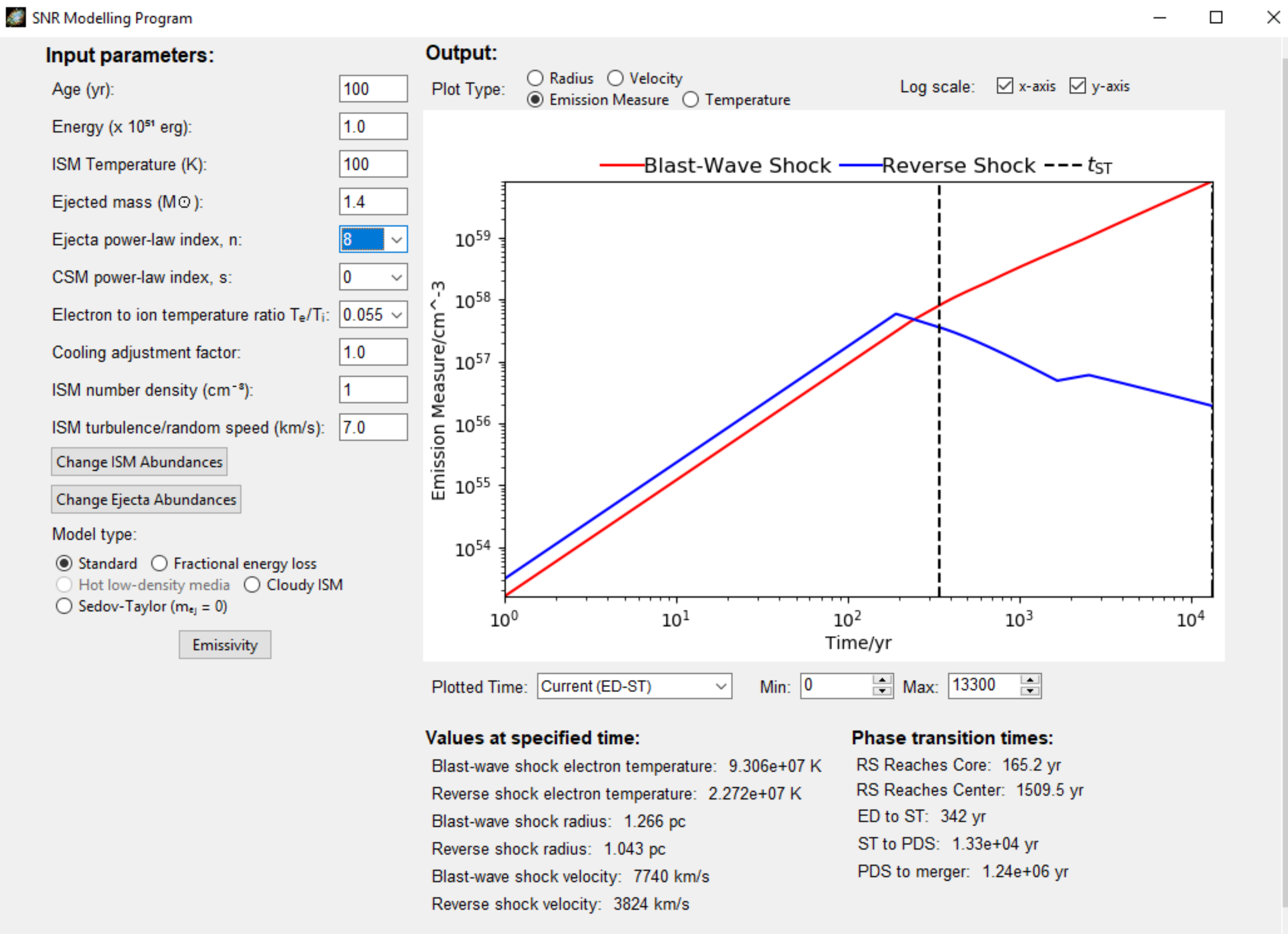}
\caption{
Temperature (top) and Emission Measure (bottom) in physical units as a function of time as calculated by the new software SNRPy. 
The case shown has s=0, n=8 and other parameters as shown in the screenshot.
 \label{figSNRPy}}
\end{figure}

The Python code SNR modelling software, SNRPy, was initially presented by LW17. 
It calculated positions of FS and RS vs. time for several values of s and n from the TM99 solutions, and for the other models
described in LW17. 
The temperatures, emission measures, interior structures and surface brightness profiles were calculated 
for s=0, n=7 and 12 using the low-resolution published CP solutions, and for low resolution WL91 solutions.
A major update to SNRPy has been made as a result of the current work. 
It now uses the new high resolution CP and WL91 solutions. 
The new code provides emission measures, interior structure and surface brightness profiles for all values of
s=0 and s=2 for all n from 6 to 14 for the self-similar CP phase.
For the unified evolution, SNRPy now provides plots of $EM$ and $EM$-weighted $T$ for  
shocked ISM and shocked ejecta as functions of time.
An example of the new calculations is shown in  Fig.~\ref{figSNRPy}.
The new SNRPy code is available for download from the website quarknova.ca in the software section, 
and from GitHub in repository denisleahy/SNRmodels.
The new CP and WL91 structure files are included in the Data directory of SNRPy. 
The animations of the PLUTO hydro simulations are provided as zip files with SNRPy. 

\acknowledgments

This work was supported by a grant from the Natural Sciences and Engineering Research Council of Canada. 
DL thanks Robert Bell, who validated the CP and WL91 solutions. 
%Yuyang Wang, who ran part of the PLUTO calculations,  Bryson Lawton for updating the SNRPy code.
%The authors thank the referee for providing useful comments which improved this work.

\appendix

\section{Emission Measures (EM) and EM-weighted Temperatures for Self-similar SNR Phases}\label{sec:selfsim}

For completeness, we present dimensionless EMs (dEM) and dimensionless EM-weighted temperatures (dT) 
for the self-similar phases of SNR evolution. 
In most cases, these values have not been presented previously in the literature, in part because previous work
was primarily concerned with calculations of shock radius and velocity and not with the interior structure required to 
calculate dEM and dT. 

\subsection{Pure ST and SNR in cloudy ISM}

Following LW17, we refer to the standard ST solution, with zero ejected
mass, as "pure ST" to differentiate it from the ST phases of TM99, with non-zero ejected mass.
WL91 present self-similar models for SNR evolution in a cloudy ISM, assuming zero ejected mass.
%Thus there is no reverse shock for these models.

For simplicity we only consider their one parameter models which depend on  $C/\tau$.
Here  $C=\rho_c/\rho_0$, with $\rho_c$ is the ISM density if the clouds were uniformly dispersed in the ISM
and $\rho_0$ is the intercloud density prior to cloud evaporation. 
The evaporation timescale parameter is $\tau=t_{evap}/t$, with $t_{evap}$ the evaporation
timescale and $t$ the age of the SNR.
The WL91 case $C/\tau=0$ is the same as the pure ST solution.

The WL91 models were recalculated by solving the self-similar differential equations given in WL91,
 using a variety of differential equation solvers. The equations were solved within
both MathCad and Mathematica software packages, using fourth-order Runga-Kutta with adaptive step size,
Burlisch-Stoer method, and a hybrid solver which uses a combination of Adams and BDF (backwards differentiation
formula). 
The results were compared and all agreed to 5 digits or better. The solutions agree with the figures shown in WL91.

Results are presented here for $C/\tau$=0 (pure ST), 1, 2 and 4. 
%The solutions are available as described at the end of this paper.
Fig.~\ref{figWLstruct} shows the interior structure (pressure, density, gas velocity and gas temperature) vs.
scaled radius, $r/R_{shock}$. 
The dimensionless $cEM$ is shown vs. dimensionless impact parameter $b$.
The integrated quantities $dEM$ and $dT$ for the WL91 solutions are given in Table~\ref{tab:tabss}.

\subsection{Early ED Phase}\label{sec:ED}

For SNR with non-zero ejected mass, the evolution starts with the early ED phase.
The ejecta has a constant density core and power-law density envelope. %, with index n. 
The self-similar evolution starts at t=0 and ends when the reverse shock approaches the ejecta core (TM99).
The self-similar solutions exist for $n>5$ and were discussed by \cite{1982Chev}.
That work presented low resolution interior structure solutions for four cases s=0, n= 7 and 12 and s=2, n=7 and 12.
Here we give interior structure solutions, dEM and dT for all values of n= 6 through 14 for both n=0 and s=2. 

We calculated the self-similar solutions, labelled CP  (Chevalier-Parker) using the methods outlined in \cite{1982Chev} and \cite{1963Parker}. 
The equations were solved within both MathCad and Mathematica software packages, using different differential
equation solvers, and comparing the results to ensure consistency.
The cases s=0 and s=2, for n=6, 7, 8, 9, 10, 11, 12, 13 and 14 were computed.
% and are made available as data tables as described at the end of the paper.

 The s=0 and s=2 solutions for n=7 and n=12 were consistent with those given in \cite{1982Chev}.
Interior solutions for s=0, n=6, 8, 10 and 12 are shown in Fig.~\ref{figs0struct} for the regions from the reverse shock
to the forward shock. 
%The density just outside the forward shock drops by a factor of 4 and the density just inside the reverse shock drops by a factor of 4. 
Interior solutions for s=2, n=6, 8, 10 and 12 are shown in  Fig.~\ref{figs2struct}.

\subsection{ $EM$s and $EM$-weighted Temperatures} 

\begin{figure}
\includegraphics[angle=0,scale=0.5] {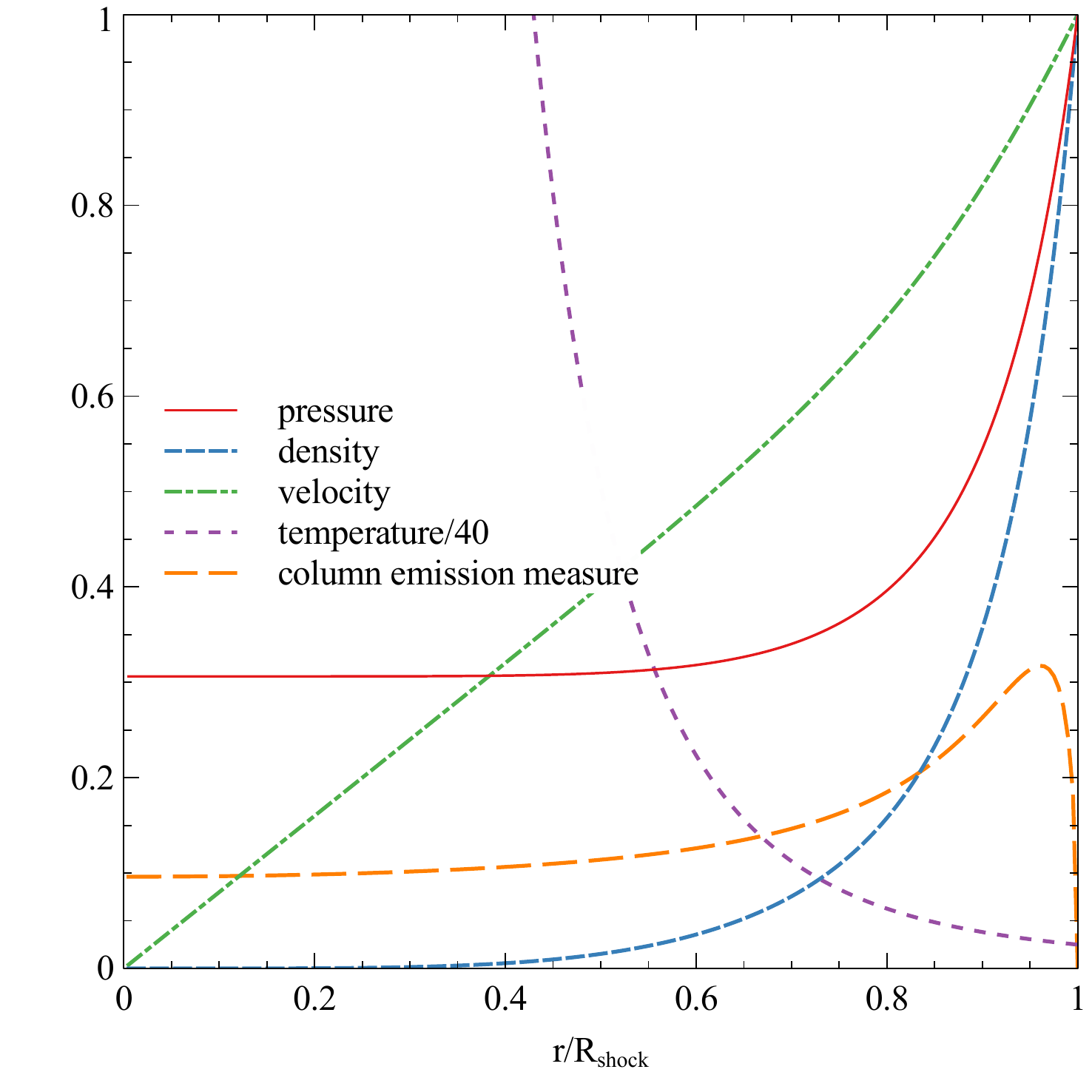}
\includegraphics[angle=0,scale=0.5] {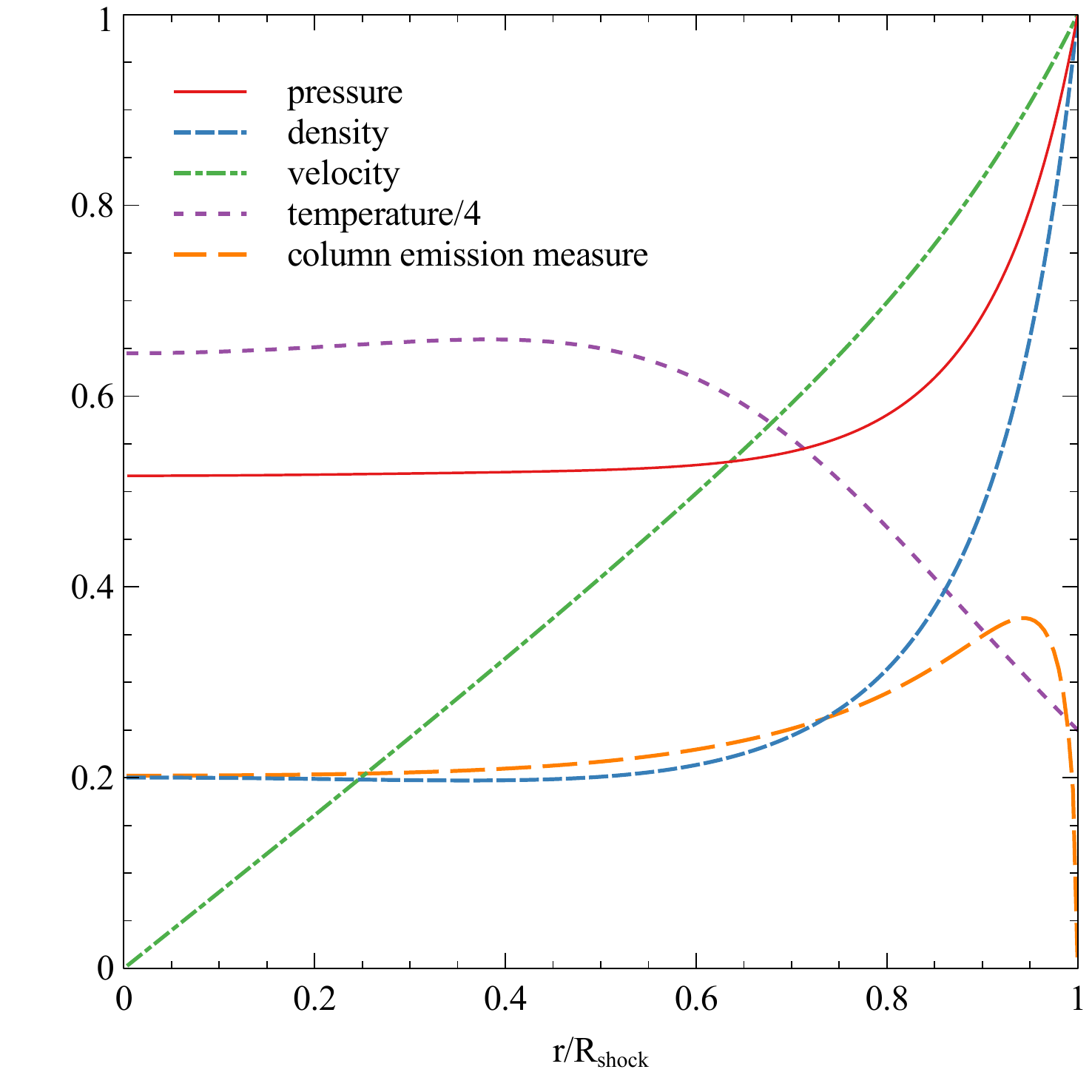}
\includegraphics[angle=0,scale=0.5] {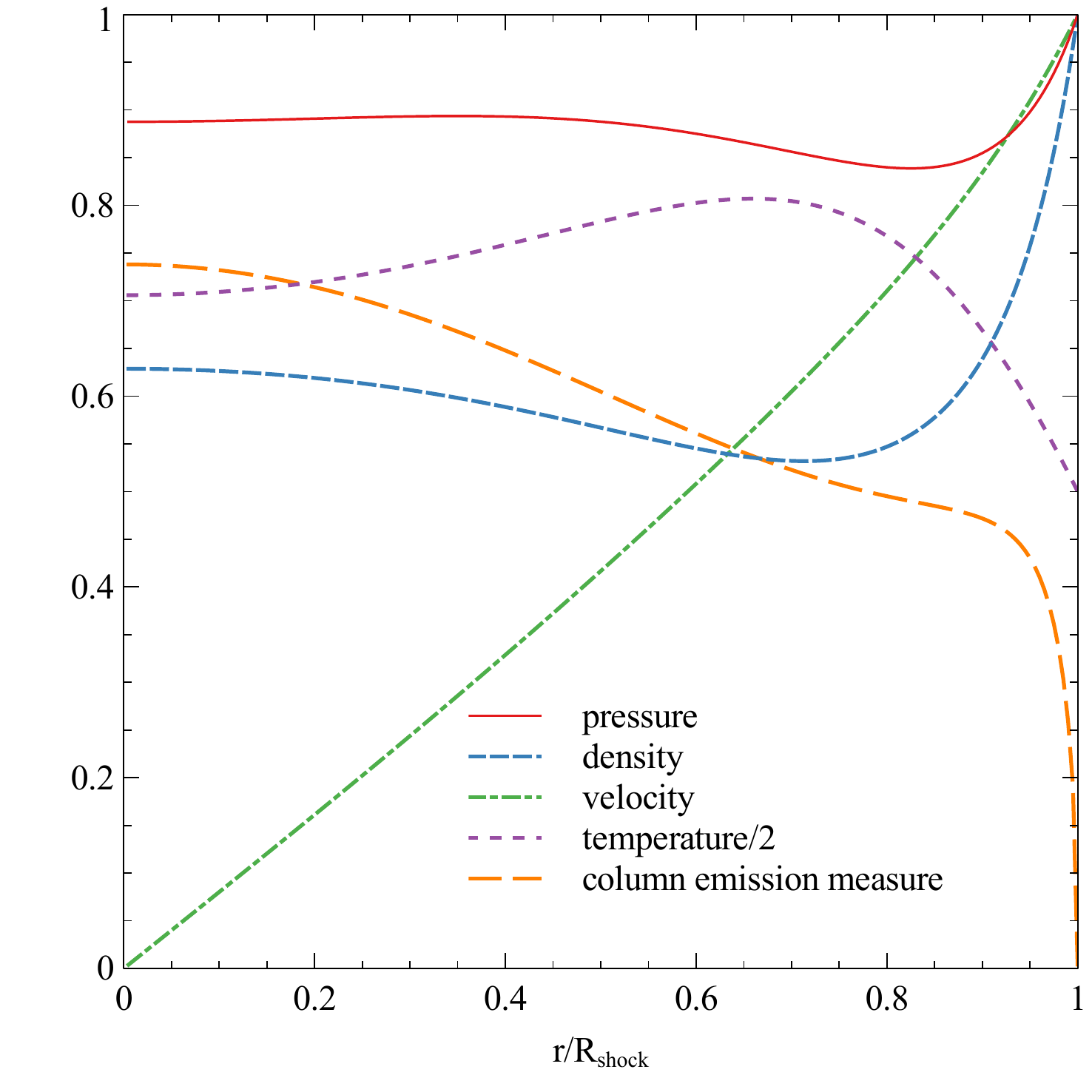}
\includegraphics[angle=0,scale=0.5] {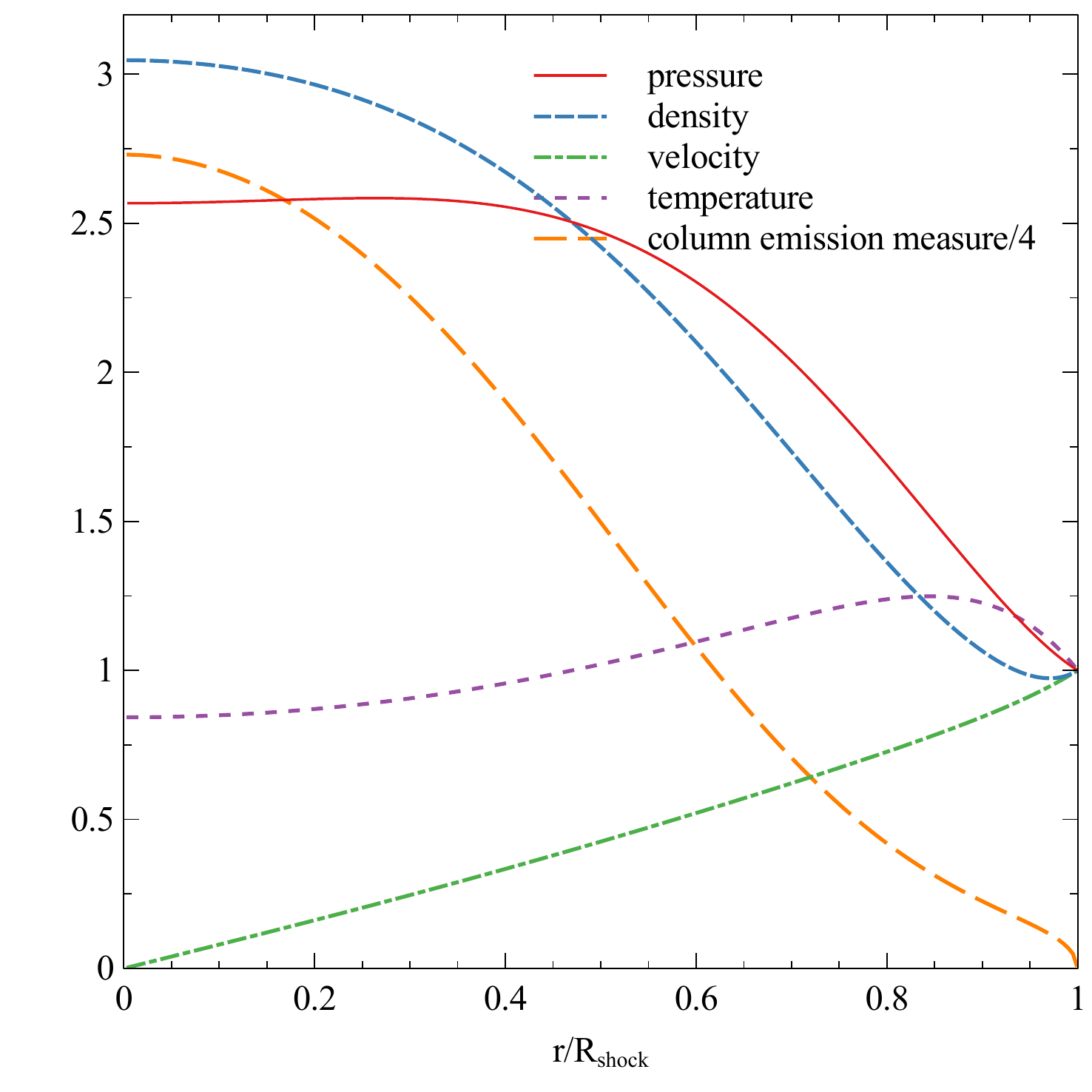}
%\plottwo{WL_CT0.pdf}{WL_CT1.pdf}
%\plottwo{WL_CT2.pdf}{WL_CT4.pdf}
\caption{
The interior structure of the WL91 self-similar solutions for $C/\tau$=0 (top left), 1 (top right), 
2 (bottom left) and 4 (bottom right). 
The pressure, density, velocity and temperature are plotted vs. radius and are scaled to their values at the forward shock. 
Further scaling factors are applied to temperature, as noted in the figure legend. 
The dimensionless column emission measure (long dash line) is plotted vs impact parameter. 
 \label{figWLstruct}}
\end{figure}

During the self-similar phases of evolution of an SNR,  $dEM_{FS}$,  $dEM_{RS}$,  $dT_{FS}$ and $dT_{RS}$ are constants;
$cEM_{FS}(b)$ and $cEM_{RS}(b)$ are functions independent of time. 
The integrated quantities $dEM$ and $dT$ for the WL solutions for $C/\tau$=0, 1, 2 and 4 are given in Table~\ref{tab:tabss}.
The dimensionless $cEM$ for the WL solutions are shown as a function of %dimensionless 
impact parameter $b=B/R_{shock}$ in  Fig.~\ref{figWLstruct}. 

\begin{figure}
\plottwo{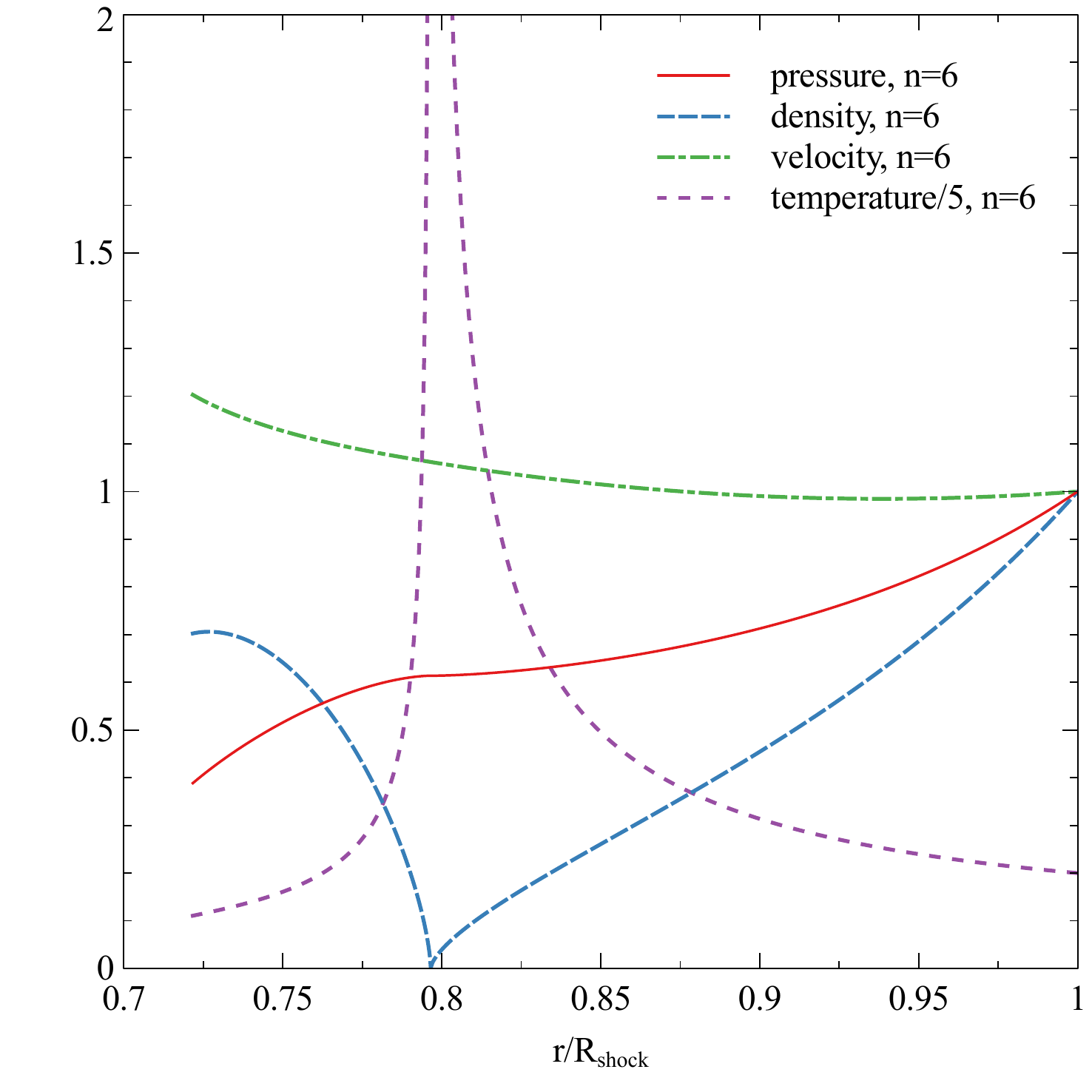}{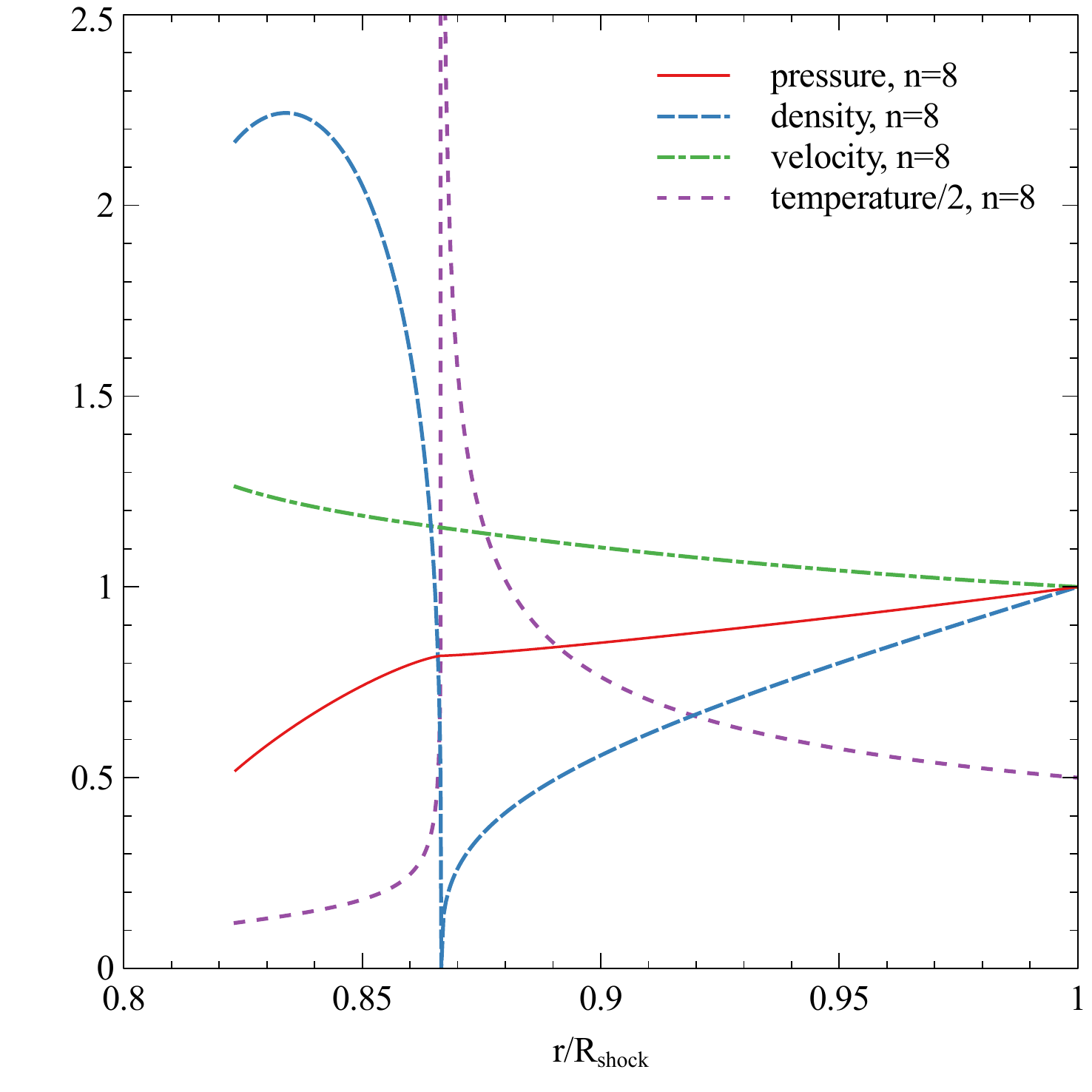}
\plottwo{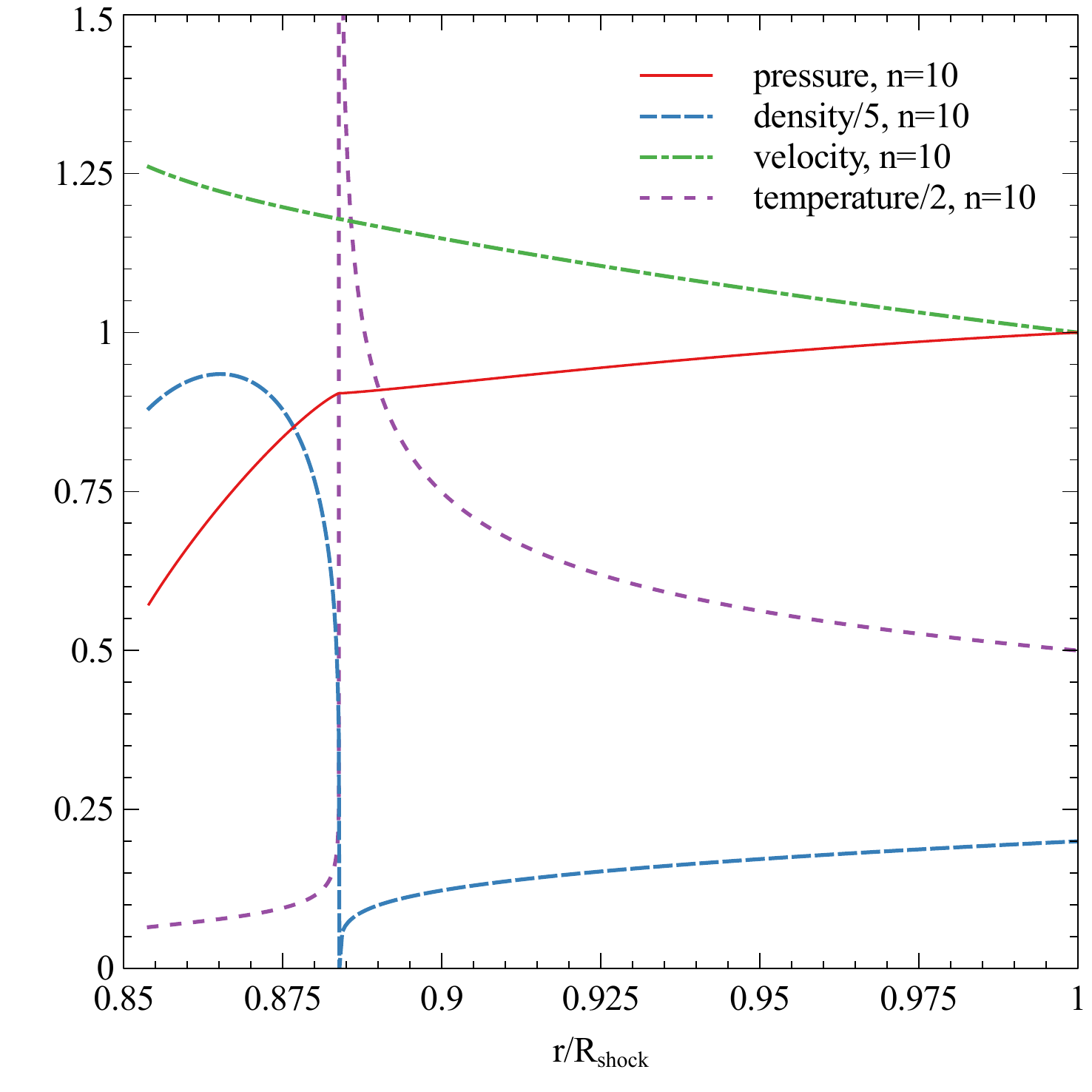}{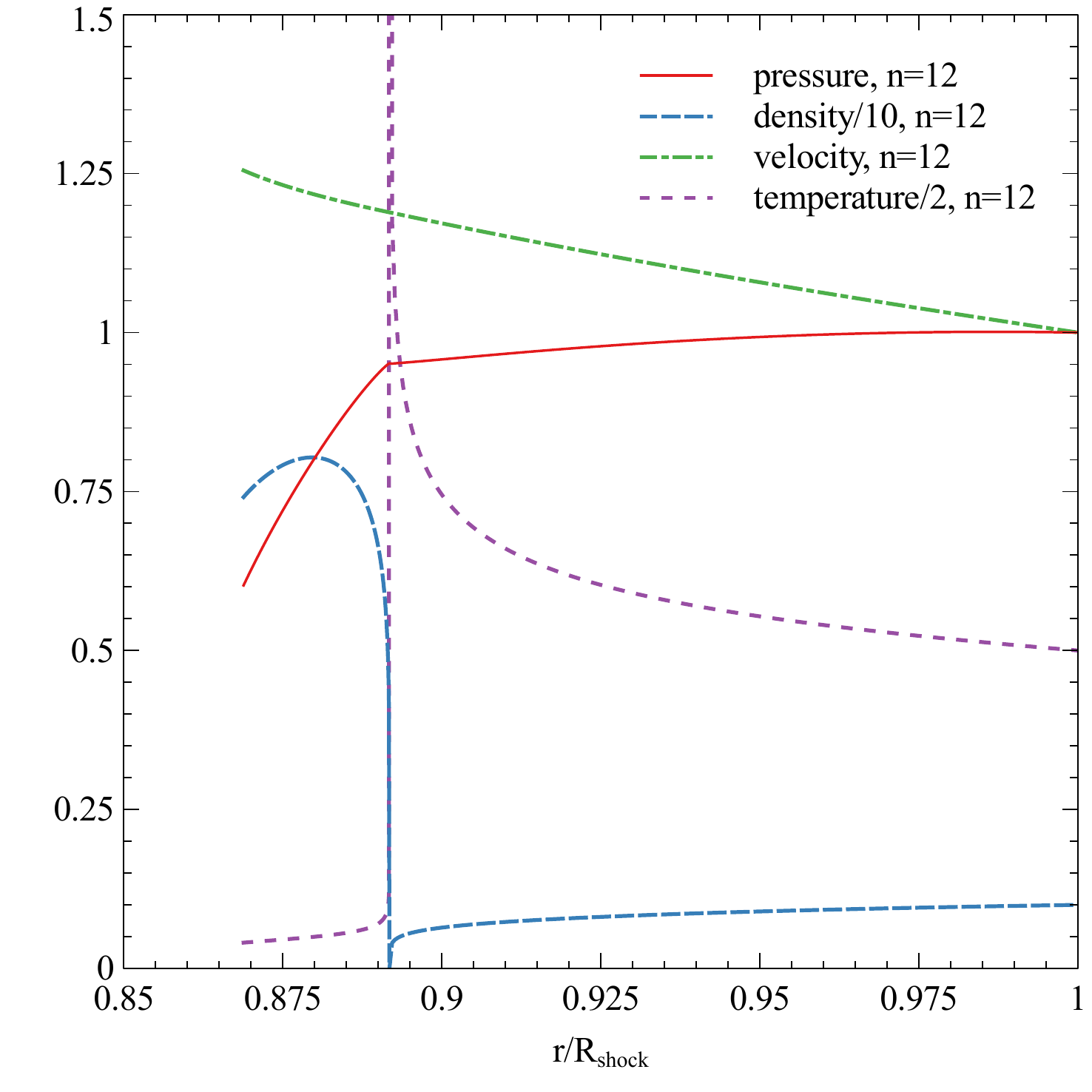}
\caption{
The interior structure of the Chevalier-Parker self-similar solutions for s=0 and n= 6 (top left), 8 (top right), 
10 (bottom left), 12 (bottom right). 
The values are scaled to the post-forward-shock values of pressure, density and temperature and the forward shock velocity. 
Further scaling factors are applied to density and temperature, as noted in the figure legend. 
The reverse shock is the point at smallest radius, and the contact discontinuity is where the density goes to  zero. 
 \label{figs0struct}}
\end{figure}

\begin{figure}
\plottwo{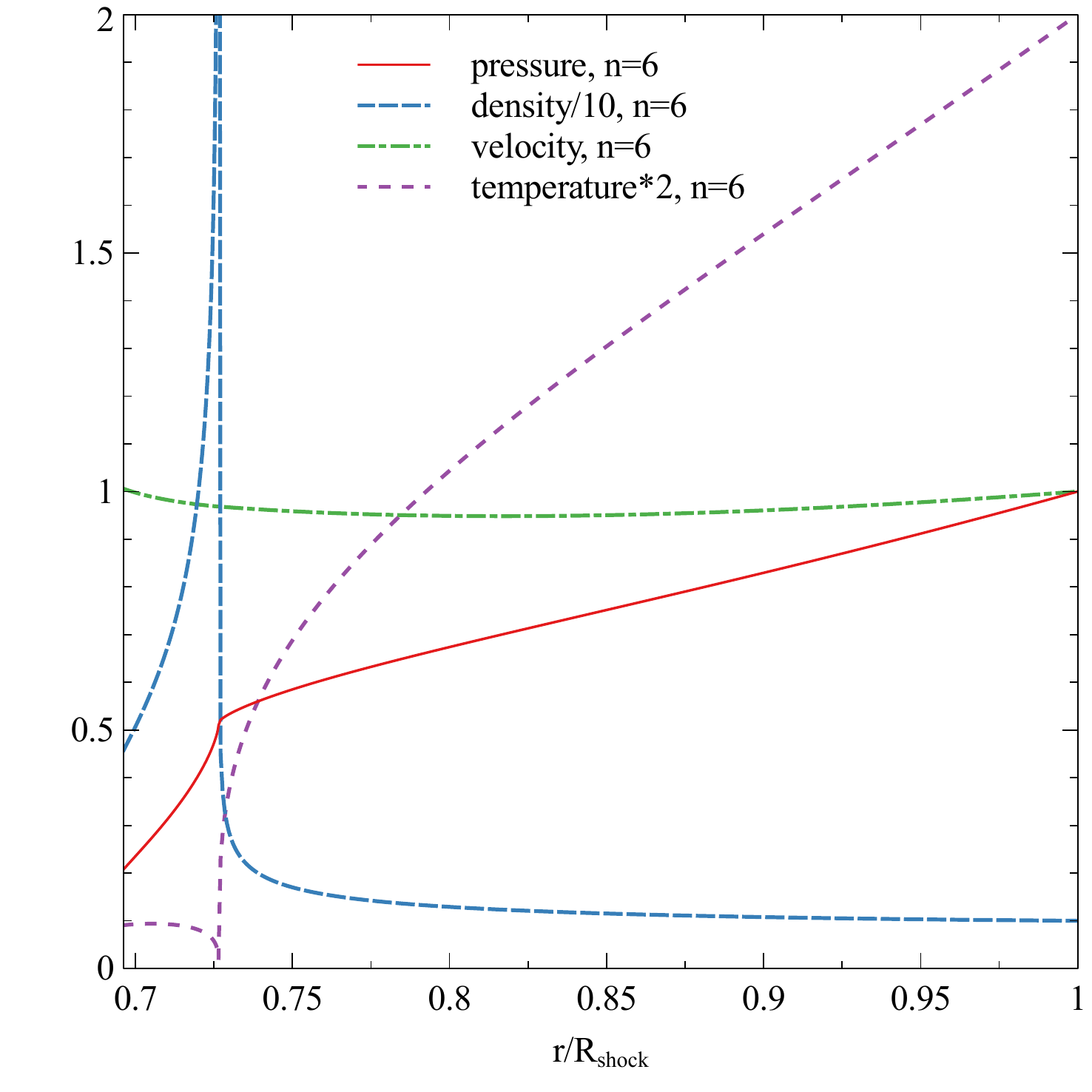}{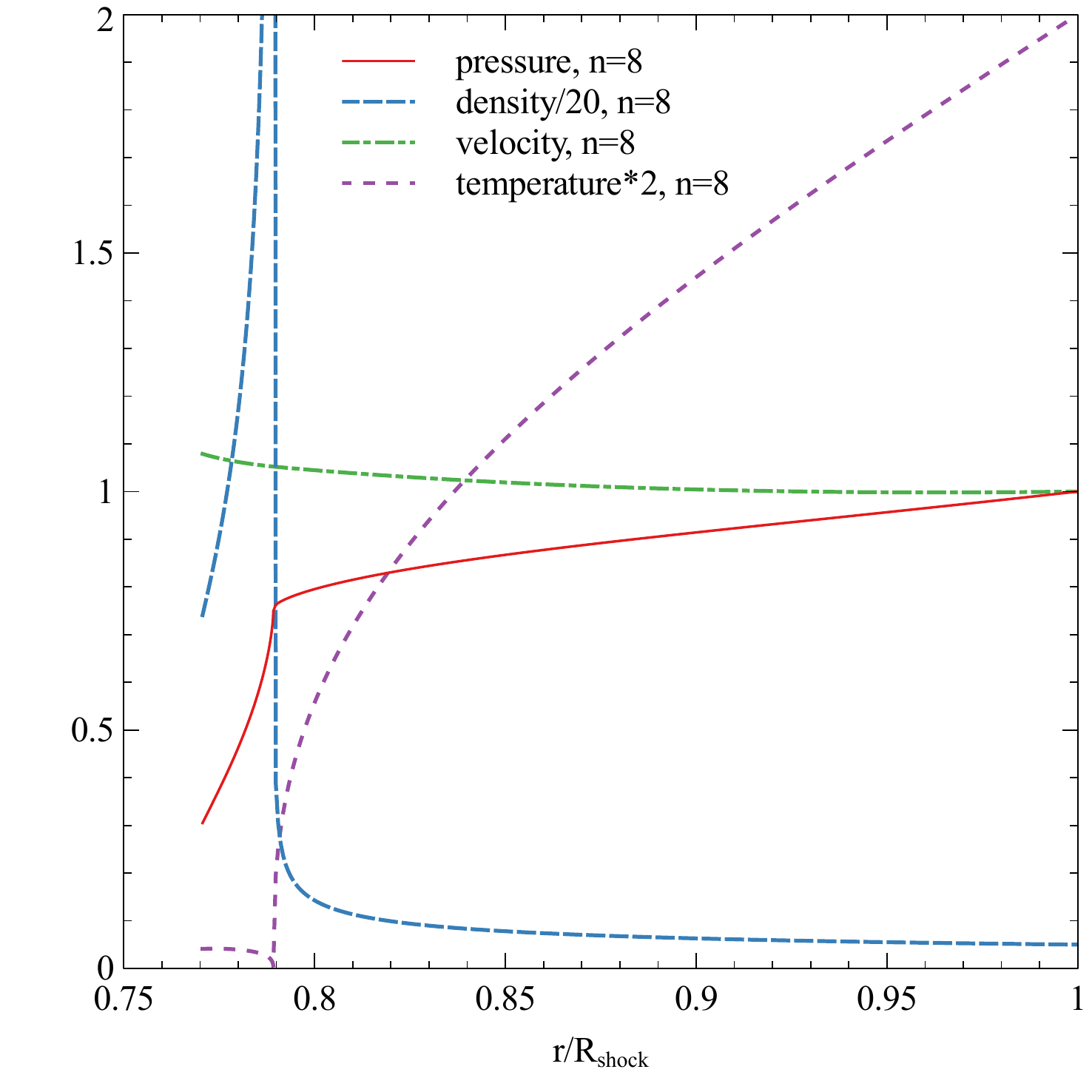}
\plottwo{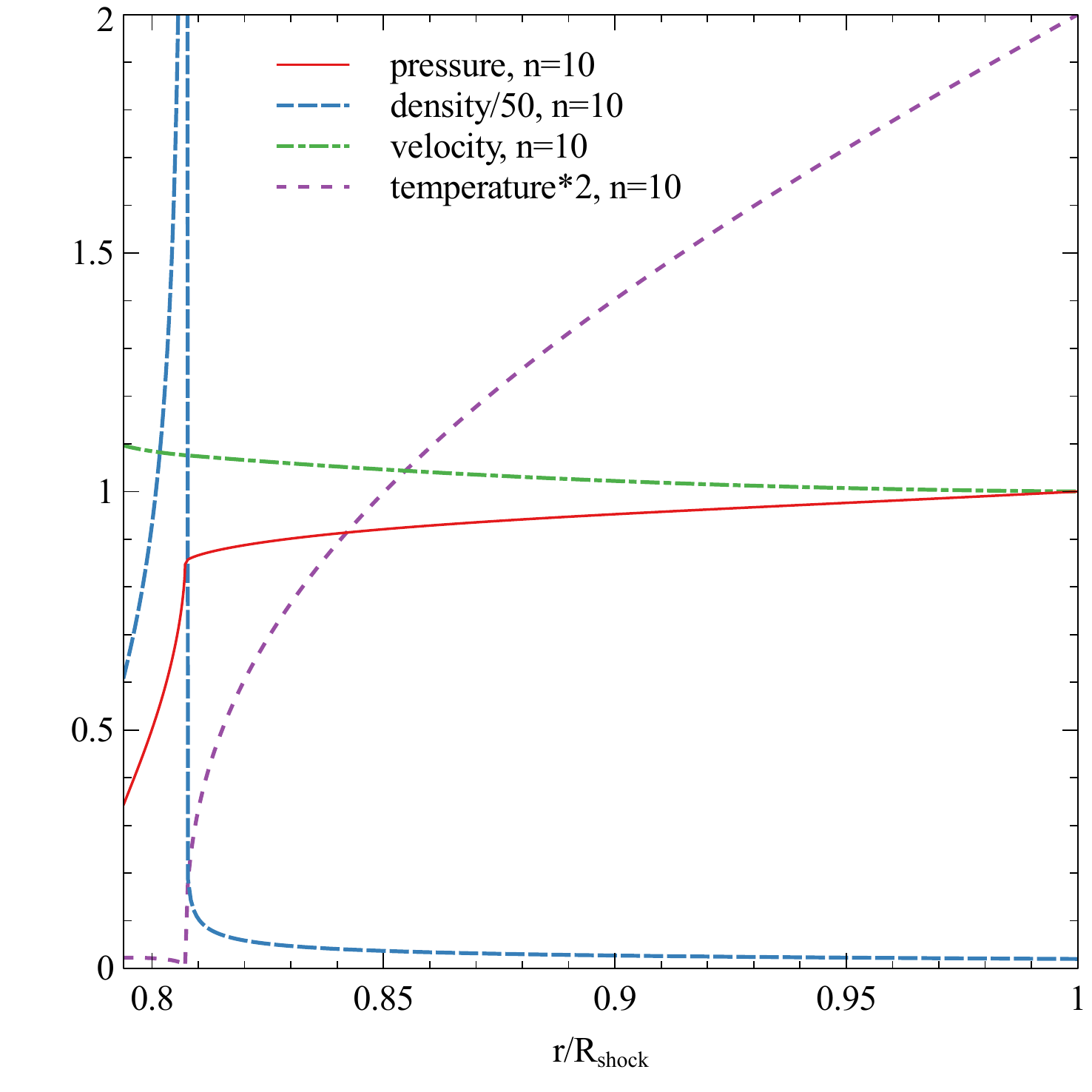}{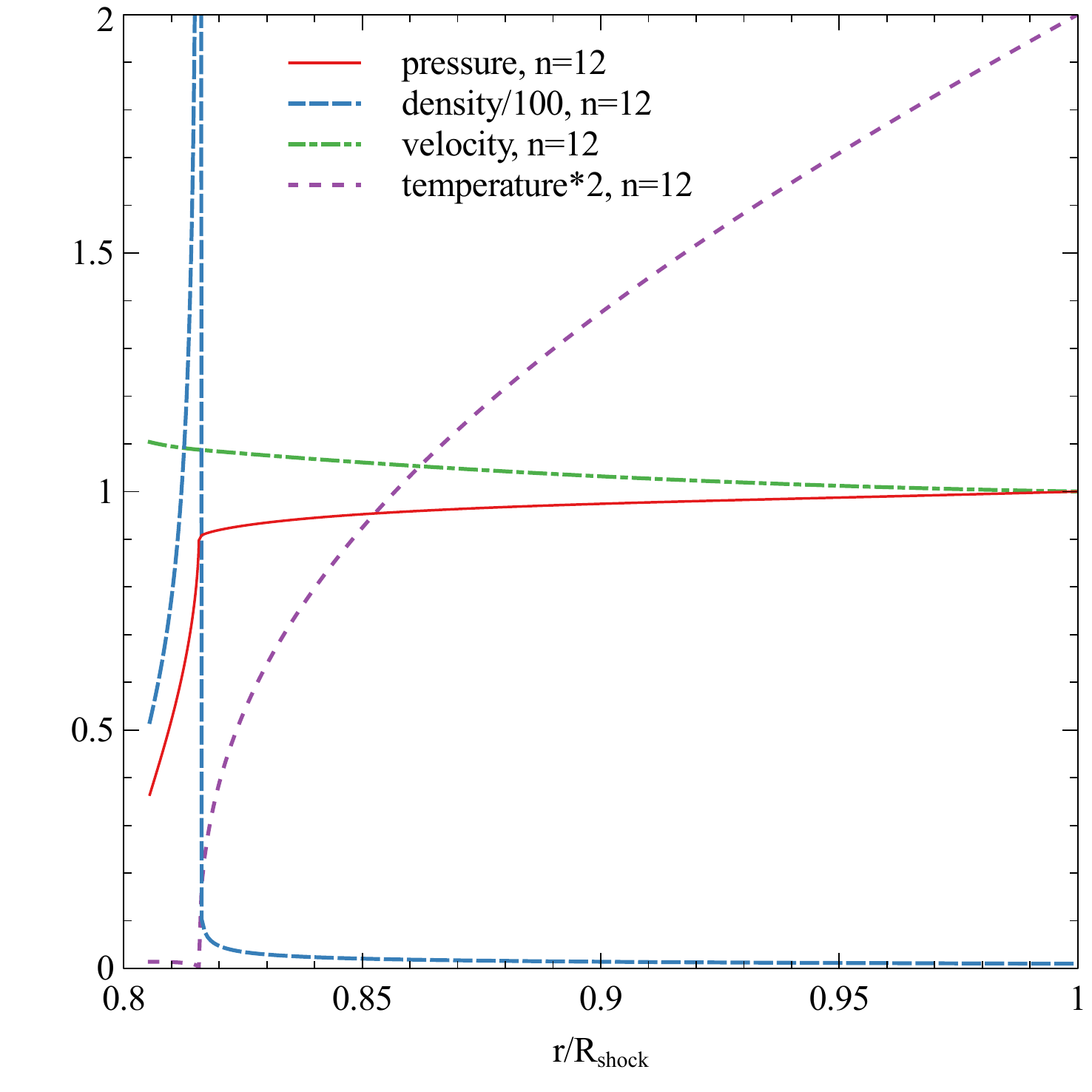}
\caption{
The interior structure of the Chevalier-Parker self-similar solutions for s=2 and n= 6 (top left), 8 (top right), 
10 (bottom left), 12 (bottom right). 
The values are scaled to the post-forward-shock values of pressure, density and temperature and the forward shock velocity. 
Further scaling factors are applied to density and temperature, as noted in the figure legend. 
The reverse shock is the point at smallest radius, and the contact discontinuity is where the density goes to  infinity.  
 \label{figs2struct}}
\end{figure}

\begin{figure}
\plottwo{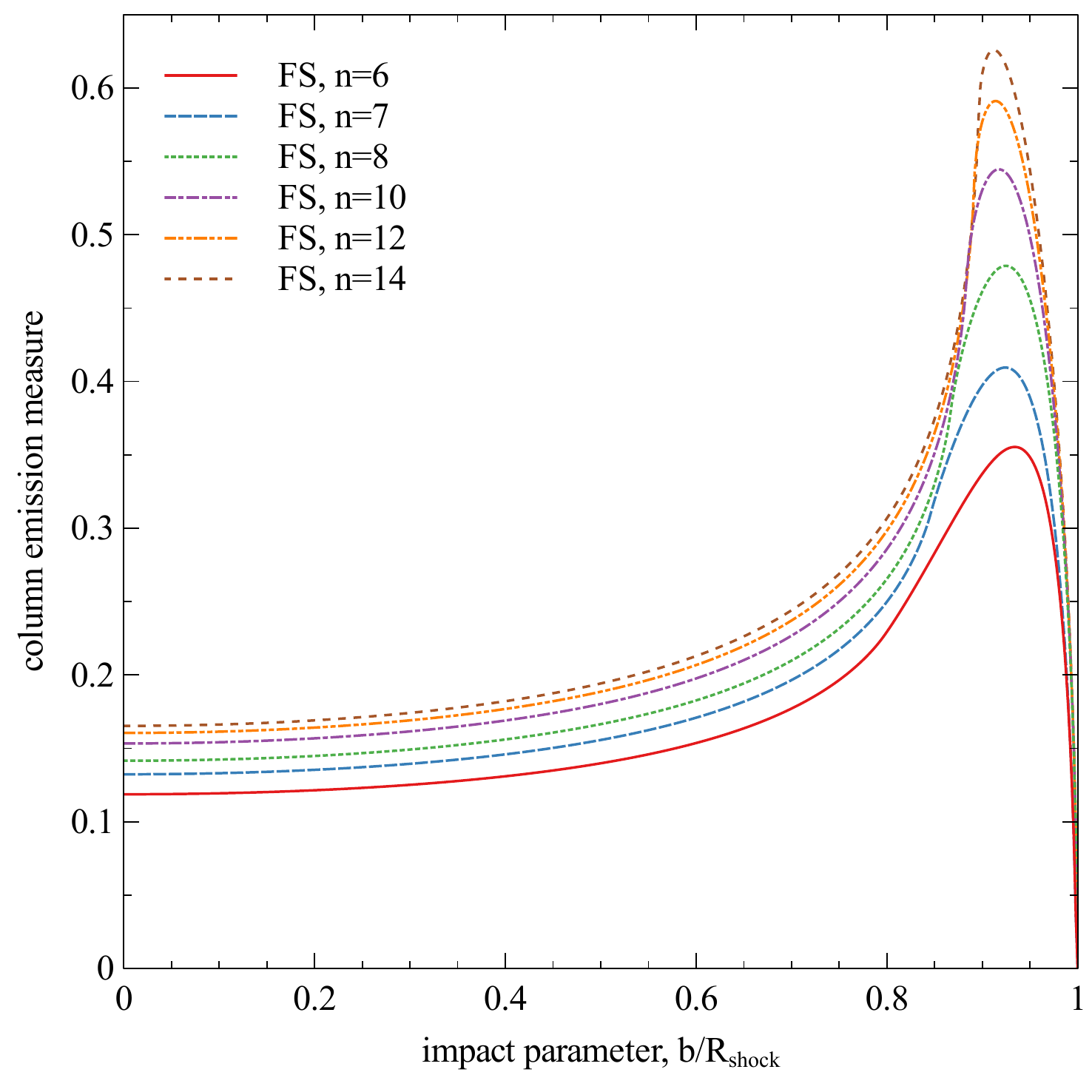}{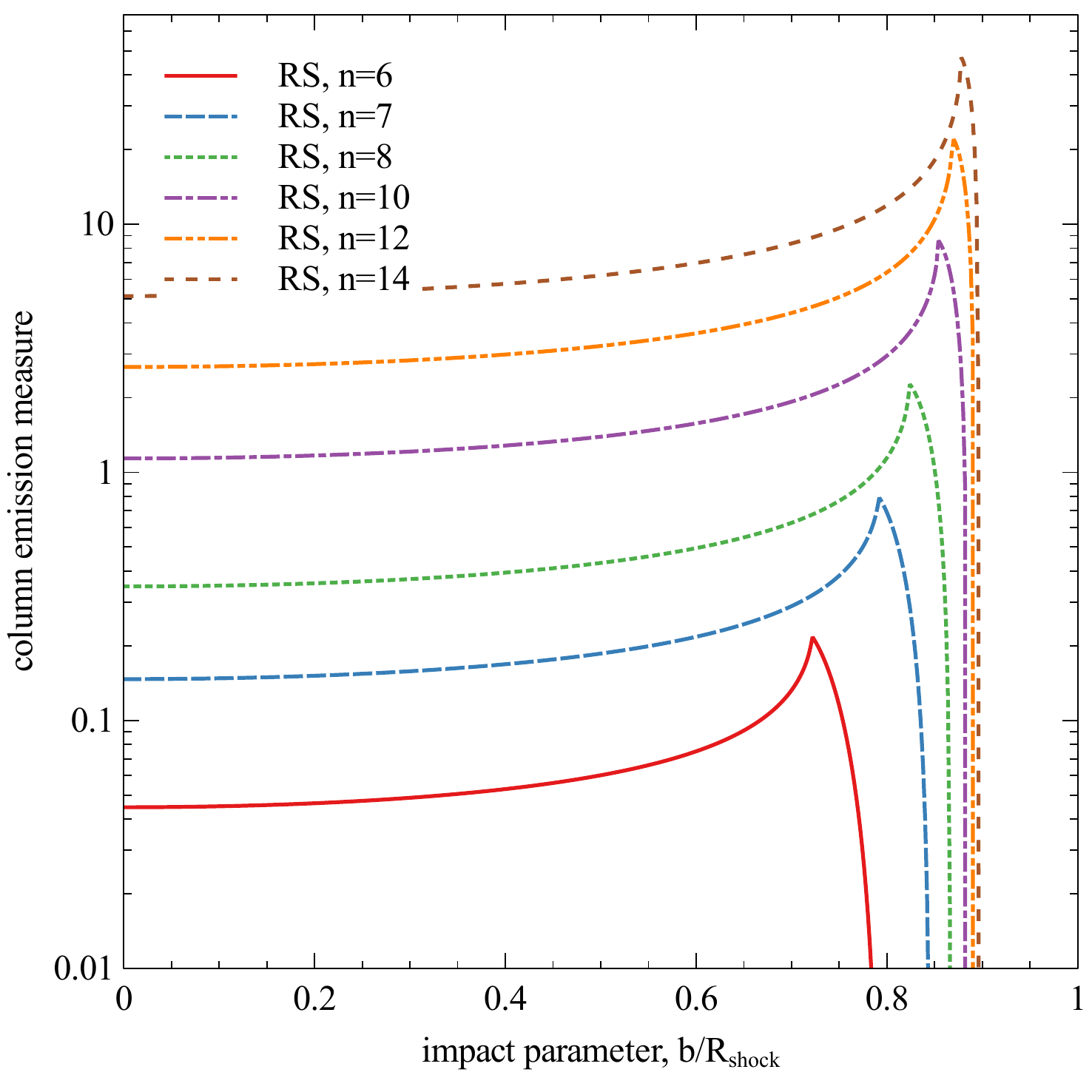}
\caption{
The dimensionless column emission measures vs. impact parameter b for the Chevalier-Parker self-similar solutions for s=0
and n= 6, 7, 8, 10, 12 and 14. 
The left panel is for material heated by the forward shock, plotted with linear scale.
The right panel is for material heated by the reverse shock, plotted with log scale.
 \label{figs0cem}}
\end{figure}

\begin{figure}
\plottwo{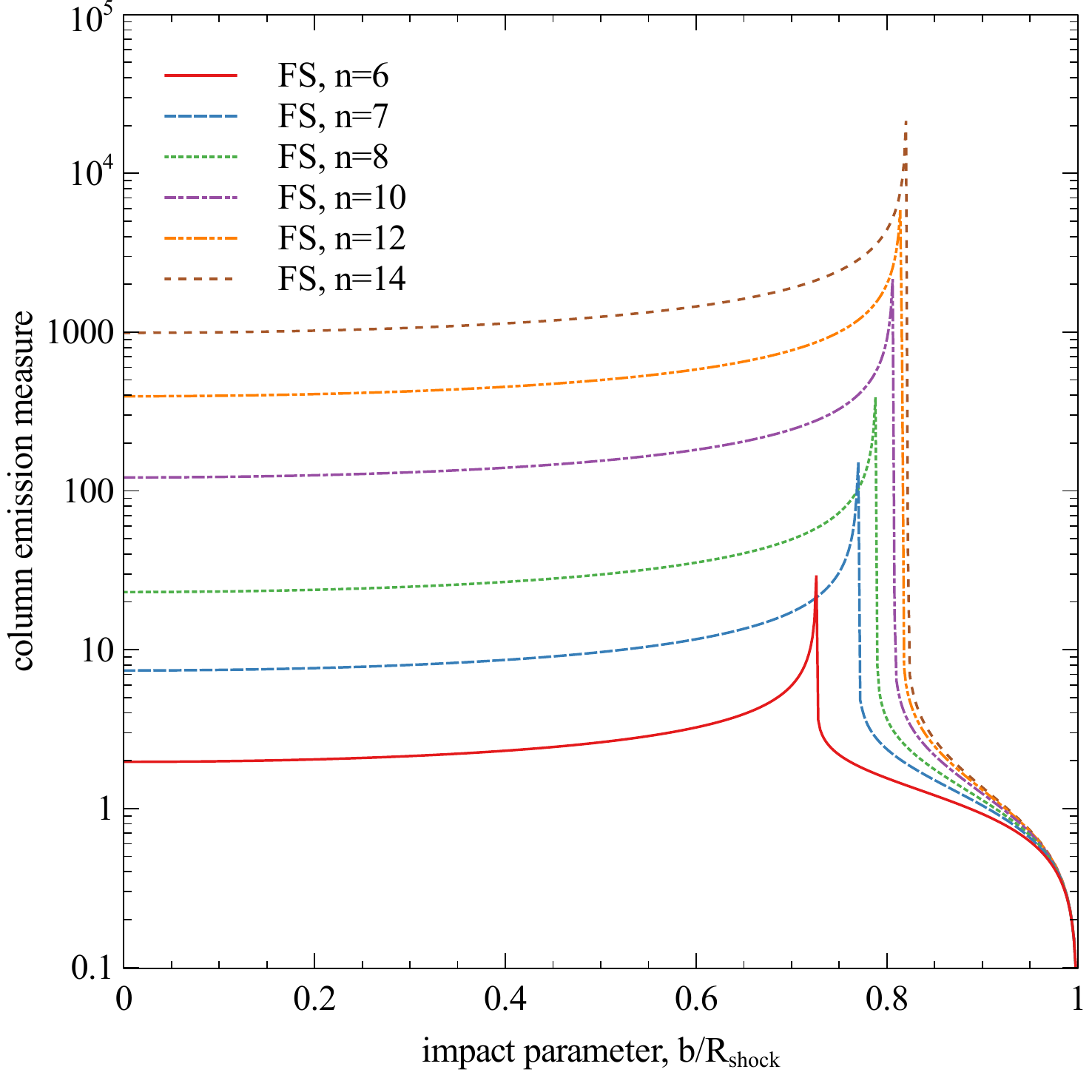}{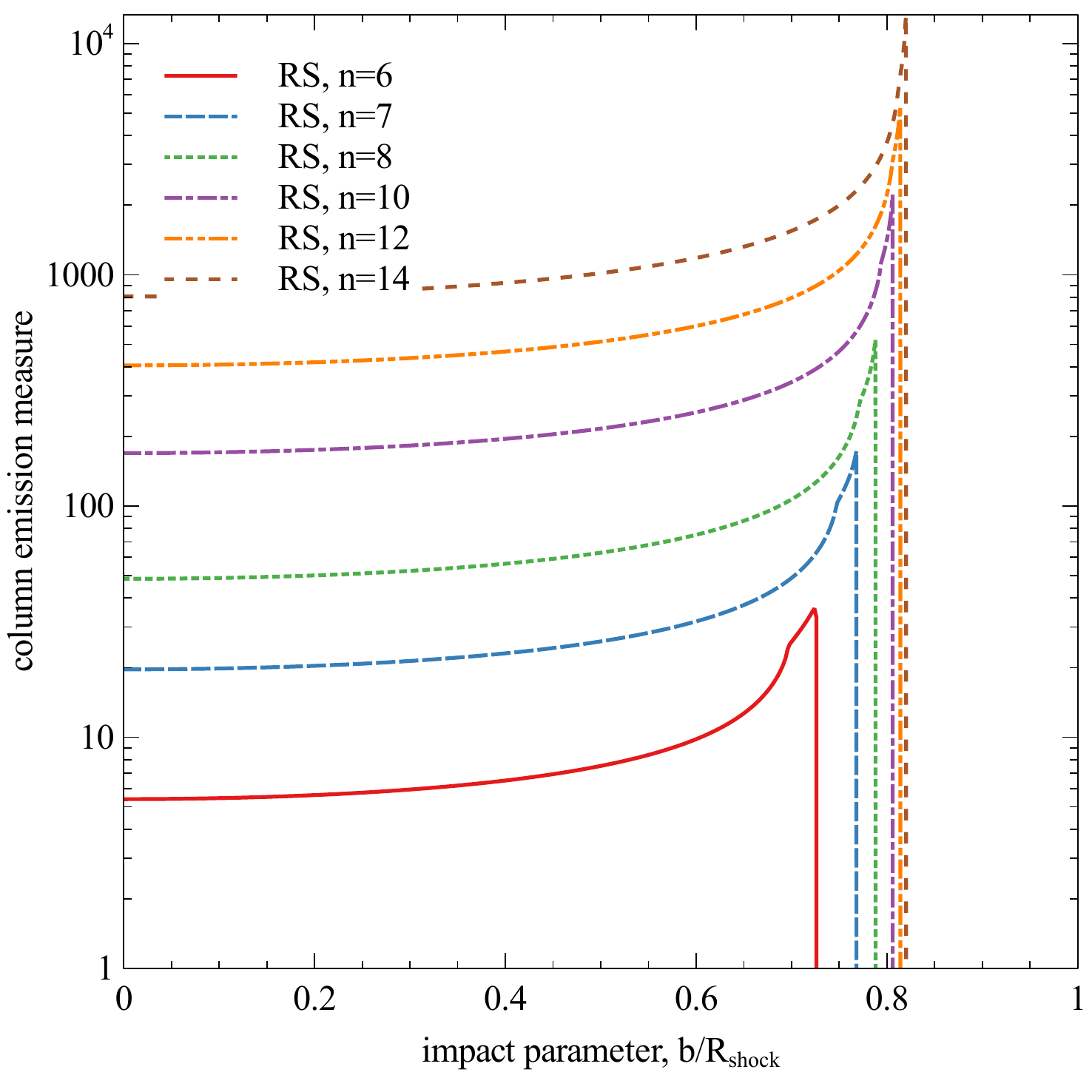}
\caption{
The dimensionless column emission measures vs. impact parameter b for the Chevalier-Parker self-similar solutions for s=2
and n= 6, 7, 8, 10, 12 and 14.  
The left panel is for material heated by the forward shock and
the right panel is for material heated by the reverse shock.
 \label{figs2cem}}
\end{figure}

$cEM$ for the CP solutions is shown as a function of %dimensionless impact parameter 
$b=B/R_{shock}$ for select s=0 cases in  Fig.~\ref{figs0cem}. 
The $cEM$ for gas between the CD and the FS is shown in the left panel. It varies smoothly with $b$, 
peaking approximately midway between the CD and the FS because of projection effects.
The $cEM$ for gas between the RS and the CD is shown in the right panel. 
Because the RS-heated gas forms a thinner and much denser shell than FS-heated gas (Fig.~\ref{figs0struct}), 
the $cEM$ is much more peaked at $b$ between the RS and the CD.
Fig.~\ref{figs2cem} shows  $cEM$ vs. $b$ for the s=2 cases.
For s=2,  both  FS-heated gas and RS-heated gas are concentrated in thin and dense shells close to the CD  (Fig.~\ref{figs2struct}). 
In projection, this explains the sharp peak in $cEM$ for both  FS-heated gas (left panel) and RS-heated gas (right panel). 
The extended tail in $cEM$ for $b$ from CD and FS is caused by projection of the low density part of the FS-heated gas.
  
 The integrated quantities $dEM$ and $dT$ from the CP solutions for FS-heated gas and RS-heated gas are given in Table~\ref{tab:tabss}.
 $dEM$ and $dT$ for FS-heated gas varies slowly with n for s=0, whereas for RS-heated gas 
 $dEM$ increases by 2 orders of magnitude and $dT$ decreases by 1 order of magnitude.
 For s=2  FS-heated gas, $dEM$ increases by a factor of 5 from n=6 to 14 and $dT$ decreases by a factor or 3.5. 
 For s=2  RS-heated gas, $dEM$ increases by  2 orders of magnitude for  n=6 to 14, and $dT$ decreases 1 order of magnitude. 
 In summary, for both s=0 and s=2  as n increases from 6 to 14,
 the RS heated gas is brighter and of lower temperature relative to  FS heated gas.

\clearpage

\clearpage

\clearpage

\begin{deluxetable}{crrrrrrrrrr}
\tabletypesize{\scriptsize}
%\rotate
\tablecaption{Coefficients for Fits to $dT_{FS}(t_s)$, $dEM_{FS}(t_s)$, $dT_{RS}(t_s)$ and $dEM_{RS}(t_s)$ \label{tab:tabhd}}%$^{a}$
\tablewidth{0pt}
\tablehead{
\colhead{} &\colhead{}   & \colhead{n=6} &  \colhead{n=7}  &  \colhead{n=8} &  \colhead{n=9} &  \colhead{n=10} &  \colhead{n=11}
  & \colhead{n=12} & \colhead{n=13} & \colhead{n=14} 
}						
\startdata
%$dT_{FS}(t_s)$ &  $dT_{FS,0}$&1.261&1.223&1.192&1.169&1.152&1.138&1.127&1.118&1.110\\
$dT_{FS}(t_s)$& $t_1$  &0.100&3.011&2.567&0.8303&0.4707&0.4149&0.3158&0.3999&0.2428 \\
&$t_2$  & 1.197&19.43&17.85&11.95&11.24&13.72&11.86&16.48&8.654 \\
&$p_1(\times 10^{-2})$ & -0.741&2.601&3.346&2.713&2.739&2.805&3.019&3.592&3.594\\
&$p_2(\times 10^{-3})$ &0.919&-5.29&-2.43&-0.352&-0.479&-0.285&-0.328&-4.50&-0.322\\
 \tableline
%$dEM_{FS}$ &    &     &  &    &   &   &   &   &   \\
% \tableline
%$dEM_{FS}(t_s)$& $dEM_{FS,0}$&0.6746&0.7542&0.8081&0.8471&0.8767&0.8998&0.9184&0.9337&0.9465\\
$dEM_{FS}(t_s)$&$t_1$  & 0.8058& 0.5343 &0.4037 &0.3352 & 0.2562 &0.2400 &0.2083 &0.2031 &0.1928 \\
&$t_2$  &  2.270&1.762&1.122&1.189&1.247&1.070&1.187&1.194 &1.149 \\
&$t_3$ &9.438&9.785&3.960&8.933&7.256&4.364&6.020&7.464&7.480 \\
&$t_4$ & 45.69&47.79&240.7&40.98&41.69&55.84&51.34&53.16&42.56\\
&$p_1$ & -0.2099&-0.2713&-0.3197&-0.3235&-0.2810&-0.2937&-0.2772&-0.2932&-0.2882\\
&$p_2(\times 10^{-2})$ & -4.777&-4.595&-11.31&-6.303&-6.753&-10.27&-8.984&-5.229&-7.038\\
&$p_3(\times 10^{-2})$ & 4.576&5.211&1.795&5.983&4.489&2.715&5.765&3.410&5.166 \\
&$p_4(\times 10^{-3})$ & $-4.81$&$-6.31$&$-7.08$&$-1.35$&$-0.056$&$-0.572$&$-9.58$&
$-0.562$&$-2.12$ \\ 
%$p_4$ & $-3.423\times 10^{-3}$&$-8.226\times 10^{-3}$&$-5.41\times 10^{-3}$&$-1.358\times 10^{-3}$&$-4.202\times 10^{-3}$&$-8.364\times 10^{-3}$&$-9.586\times 10^{-3}$&$-1.4974\times 10^{-4}$&$-1.500\times 10^{-4}$ \\ 
 \tableline
%$dT_{RS}$ &    &     &  &    &   &   &   &   &   \\
% \tableline
%$dT_{RS}(t_s)$&  $dT_{RS,0}$&0.8868&0.5204&0.3368&0.2347&0.1723&0.1318&0.1039&0.084&0.0693\\
$dT_{RS}(t_s)$& $t_1$  &0.1001&0.2290&0.4398&0.4016&0.3075&0.2618&0.2382&0.2162&0.1966 \\
&$t_2$  &1.376&1.229&2.037&2.523&2.686&2.492&2.766&2.801&2.925 \\
&$t_3$  &5.023 &6.995 &4.631 &4.504 &4.520&4.582& 4.558&4.508&4.457\\
&$p_1$ & -0.0392&0.0782&0.5480&0.6564&0.6537&0.6436&0.6964&0.7265&0.7601\\
&$p_2$ & 1.264&1.166&1.562&1.842&1.996&1.841&2.094&2.079&2.247\\
&$p_3$ & 0.7157&0.7137&0.7131&0.7190&0.7159&0.7204&0.7209&0.7211&0.7242 \\
 \tableline
%$dEM_{RS}$ &    &     &  &    &   &   &   &   &   \\
% \tableline
%$dEM_{RS}(t_s)$& $dEM_{RS,0}$&0.1604&0.6250&1.549&3.087&5.400&8.634&12.98&18.53&25.50\\
$dEM_{RS}(t_s)$&$t_1$  & 0.6146&0.4272&0.3382&0.3221&0.2357&0.2209&0.2104&0.1939&0.1747 \\
&$t_2$  &  2.833&2.657&2.953&2.462&2.799&2.832&2.665&2.730&2.681 \\
&$t_3$ &4.482&4.888&4.513&4.984&4.819&4.861&4.922&4.821&4.848 \\
&$p_1$ &2.465&2.652&2.733&2.955&2.718&2.781&2.884&2.891&2.882\\
&$p_2$ & 1.247&1.384&0.781&1.335&1.008&1.023&0.989&1.009&1.006\\
&$p_3$ & 1.919&1.919&1.919&1.920&1.926&1.927&1.941&1.928&1.933 \\
% \tableline
\enddata
%\tablenotetext{a}{All models have SN ejecta mass of 1.4$M_{\odot}$. }
\end{deluxetable}

\clearpage

\begin{deluxetable}{ccccrrrrrr}
\tabletypesize{\scriptsize}
%\rotate
\tablecaption{Historical SNR Observed Quantities \label{tab:snrdata}} 
\tablewidth{0pt}
\tablehead{
\colhead{SNR}  & \colhead{Age$^a$}  & \colhead{Type}  & \colhead{Distance}   &  \colhead{Radius$^{b}$}      & \colhead {EM$^{b,c}$}     & \colhead {kT$^{c}$} & Refs$^{d}$ \\ %& \colhead {$EM_2^{c}$}     & \colhead {$kT_2^b}$} \\
\colhead{}    & \colhead{(yr)}  & \colhead{}           & \colhead{(kpc)}     & \colhead{(pc)}      & \colhead {($10^{58}$cm$^{-3}$)} & \colhead {(keV)} & \\ %& \colhead {($10^{58}$cm$^{-3}$)} & \colhead {(keV)}\\
% & \colhead{(yr)} 
}
\startdata
 G$4.5+6.8$ (Kepler)    & 407        & Ia & $ 5.1^{+0.8}_{-0.7}$ & $ 2.64$d$_{5.1} $ & $0.6007^{+0.0125}_{-0.0202}$d$_{5.1}^2$  & $1.06 \pm 0.03 $  & 1, 2, 3 \\ 					
  G$120.1+1.4$ (Tycho)  & $ 434 $ & Ia & $ 4.0\pm1.0$            & 4.96d$_{4}$    & $3.421^{+0.144}_{-0.025}$d$_{4}^2$              & $ 0.41 \pm 0.01 $ &  4, 5, 3 \\  
 G$327.6+14.6$ (SN1006) & $ 1002 $        & Ia & $ 2.18 \pm 0.08 $   &  9.50d$_{2.18}$    & $0.0485$d$_{2.18}^2$                & $ 0.4 \pm 0.1 $   &  6, 7, 8 \\   
\enddata
\label{tab:TBLobserved}
\tablenotetext{a} {The ages are taken, from the year the SN event occurred to the year of observation for the integrated X-Ray Spectra of the SNRs (need to check obs dates).}
\tablenotetext{b} {The radii of the SNRs are obtained from the X-ray observations. $d_{5.1}$ is distance in units of 5.1 kpc, $d_{4}$  is distance in units of 4 kpc, and $d_{2.18}$  is distance in units of 2.18 kpc.}
\tablenotetext{c}{$EM$ and $kT$ are for Shocked ISM.}
\tablenotetext{d}{References are for the SNe type, distances and EM \& kT values, respectively.}
\tablerefs {(1) \cite{1999Kinugasa}, (2) \cite{2016Sankrit}, (3) \cite{2015Katsuda}, (4) \cite{2008Krause}, (5) \cite{2010Hayato}, (6) \cite{1996Schaefer}, (7) \cite{2003Winkler}, (8) \cite{2013Uchida} }
%\tablenotetext{b}{For SNRs with only one measured thermal plasma component, $EM$ and $kT$ are given; for SNRs with two measured thermal plasma components,
%$EM$ and $kT$ are given for the first component and $EM_2$ and $kT_2$ are given for the second component.}
\end{deluxetable}

%\clearpage

\begin{deluxetable}{crrrrrrrrrr}
\tabletypesize{\scriptsize}
%\rotate
\tablecaption{Historical SNR Shocked Ejecta Quantities \label{tab:snrdata2}} %
\tablewidth{0pt}
\tablehead{
\colhead{SNR }   & \colhead {$EM_1^{a}$ }     & \colhead {$kT_1 $} & \colhead {$EM_2^{a} $}     & \colhead {$kT_2 $}  & \colhead {$EM_3^{a} $}     & \colhead {$kT_3 $} \\
\colhead{}        & \colhead {($10^{58}$cm$^{-3}$)} & \colhead {(keV)} & \colhead {($10^{58}$cm$^{-3}$)} & \colhead {(keV)}
& \colhead {($10^{58}$cm$^{-3}$)} & \colhead {(keV)} 
}
\startdata
Kepler     & $ 2.990^{+0.355}_{-0.031} \times 10^{-5}$  &  $0.37 \pm 0.01$   & $2.723^{+0.015}_{-0.013} \times 10^{-5}$   & $ 2.08^{+0.01}_{-0.02}$    &  $18.07^{+0.22}_{-0.19} \times 10^{-5}$    &  $2.59 \pm 0.01$ \\   
Tycho   & $ 17.25^{+0.88}_{-1.48} \times 10^{-5}$  &  $0.70^{+0.02}_{-0.01}$   & $ 4.353 \pm 0.004 \times 10^{-7}$   & $ 0.96 \pm 0.01$ &  $2.06 \pm 0.2  \times 10^{-8} $  &  $9.34^{+0.03}_{-0.25}$ \\  
SN1006  & $ 2.00 \times 10^{-6} $  &  $0.48 \pm 0.01$   & $ 5.91 \times 10^{-6}$   & $ 1.73 \pm 0.03 $    &  \nodata    &  \nodata \\     
\enddata
\label{tab:TBLobserved}
\tablenotetext{a}{The $EM_1$, $EM_2$ and $EM_3$ values scale with distance as $d_{5.1}^2$ for Kepler, as $d_{4}^2$ for Tycho and 
as $d_{2.18}^2$ for SN1006.}
\end{deluxetable}

\clearpage

\begin{deluxetable}{crrrrrrrrcrrrr}
\tabletypesize{\scriptsize}
%\rotate
\tablecaption{Historical SNR Model Results \label{tab:snrmodels}} %$^{a}$
\tablewidth{0pt}
\tablehead{
\colhead{SNR} & \colhead {Dist} 
& \colhead {s} & \colhead {n} & \colhead {$M_{ej}$} & \colhead{Age} &  \colhead{Energy}    & \colhead{$n_0$(s=0)}    & \colhead{$\rho_s$(s=2)}  & \colhead {$\textrm{EM}_{\textrm{RS}} $} & \colhead {$\textrm{kT}_{\textrm{RS}} $}  & \colhead {$\textrm{R}_{\textrm{RS}} $}  \\
\colhead{}   & \colhead{(kpc)}  & \colhead{}   &   & \colhead{($M_{\odot}$)} & \colhead{(yr)}  & \colhead{($10^{51}$erg)}  & \colhead{(cm$^{-3}$)} & \colhead{(M$_{\odot}$s/(km~yr))} & \colhead {($10^{58}$cm$^{-3}$)} & \colhead {(keV)} & \colhead{(pc)}\\
% & \colhead {Sedov Age}    % & \colhead{(yr)} 
}
\startdata
G$4.5 +6.8$   &$5.1$ & 0 & 7 & 1.2 &1550&0.0374&0.893 & n/a                   &$1.25 \times 10^{-4}$ & 1.92 & 2.04 \\   
(Kepler)      &      & 2 & 6 & 1.2 &332 &0.718 &n/a   & $4.36 \times 10^{-8}$ &$2.43\times 10^{-4}$ & 3.63 & 1.84 \\  
 	  	      &      & 2 & 7 & 1.2 &103 &5.26  &n/a   & $1.84 \times 10^{-8}$ &$1.59\times 10^{-4}$ & 16.4 & 1.97 \\                                           
              &      & 2 & 8 & 1.2 &157 &2.71   &n/a   & $2.43 \times 10^{-8}$ &$6.72 \times 10^{-4}$ & 5.69 & 2.04 \\                                          
              &      & 2 & 6 & 1.0 &332 &0.676 &n/a   & $4.36 \times 10^{-8}$ &$2.43 \times 10^{-4}$ & 3.63 & 1.84 \\ 
              &      & 2 & 6 & 1.4 &332 &0.756 &n/a   & $4.36 \times 10^{-8}$ &$2.43 \times 10^{-4}$ & 3.63 &  1.84  \\ 
					 
		      &$4.4$ & 2 & 6 & 1.2 &282 &0.582 &n/a   & $3.51 \times 10^{-7}$ &$1.81 \times 10^{-4}$ & 3.70 & 1.59   \\	

              &$5.9$ & 2 & 6 & 1.2 &392 &0.886&n/a   & $5.46 \times 10^{-8}$ &$3.25 \times 10^{-4}$ & 3.55 & 2.13  \\                                           
                   
G$120.1 +1.4$ &$4.0$ & 0 & 7 & 1.2 &4050&0.0695&0.967 & n/a                   &$3.10 \times 10^{-5}$ & 1.24 & 1.32 \\
(Tycho)       & 	 & 2 & 6 & 1.2 &1830&0.281 &n/a	  & $1.43 \times 10^{-7}$ &$1.38 \times 10^{-3}$ & 0.59 & 3.45 \\
		      & 	 & 2 & 7 & 1.2 &378 &3.41  &n/a	  & $6.03 \times 10^{-8}$ &$9.05 \times 10^{-4}$ & 5.41 & 3.71 \\
			  & 	 & 2 & 8 & 1.2 &688 &1.03  &n/a	  & $7.96 \times 10^{-8}$ &$3.83 \times 10^{-3}$ & 1.52 & 3.82 \\
			  & 	 & 2 & 7 & 1.0 &378 &3.11  &n/a	  & $6.03\times 10^{-8}$ &$9.05 \times 10^{-4}$ & 5.41 & 3.71 \\	
			  & 	 & 2 & 7 & 1.4 &378 &3.68  &n/a   & $5.84 \times 10^{-8}$ &$9.05 \times 10^{-4}$ & 5.41 & 3.71 \\				  
			  &$3.0$ & 2 & 7 & 1.2 &278 &2.47  &n/a   & $3.92 \times 10^{-8}$ &$5.09 \times 10^{-4}$ & 5.50  & 2.78 \\	
			&$3.0$ & 2 & 8 & 1.2 &498 &0.825  &n/a   & $5.16 \times 10^{-8}$ &$2.15 \times 10^{-3}$ & 1.58  & 2.86 \\
			  &$5.0$ & 2 & 7 & 1.2 &480 &4.37  &n/a   & $8.43 \times 10^{-8}$ &$1.42 \times 10^{-3}$ & 5.33 & 4.63 \\
%			  &$4.5$ & 2 & 7 & 1.2 &428 &3.86  &n/a   & $7.19 \times 10^{-8}$ &$1.15 \times 10^{-3}$ & 5.37 & 4.17 \\					 			 
G$327.6 +14.6$&$2.18$& 0 & 7 & 1.2 &8460&0.0237&0.0394& n/a	                  &$3.61 \times 10^{-7}$ & 0.742& 6.66  \\
(SN1006)	  &      & 2 & 7 & 1.2 &604 &2.75  &n/a	  & $9.94 \times 10^{-9}$ &$1.28 \times 10^{-5}$ & 6.20  & 7.10   \\	
			  & 	 & 2 & 8 & 1.2 &924 &1.32  &n/a   & $1.31 \times 10^{-8}$ &$5.43 \times 10^{-5}$ & 2.14 & 7.32  \\
			  & 	 & 2 & 9 & 1.2 &1140&0.863 &n/a	  & $1.46 \times 10^{-8}$ &$1.35 \times 10^{-4}$ & 1.13& 7.45  \\
			  & 	 & 2 & 8 & 1.0 &924 &1.18 &n/a	  & $1.31 \times 10^{-8}$ &$5.43 \times 10^{-5}$ & 2.14 & 7.32  \\
		      & 	 & 2 & 8 & 1.4 &924 &1.45 &n/a	  & $1.31 \times 10^{-8}$ &$5.43 \times 10^{-5}$ & 2.14 & 7.32  \\			 	
    		   	&$2.10$& 2 & 9 & 1.2 &1095 &0.84  &n/a	  & $1.38 \times 10^{-8}$ &$1.23 \times 10^{-4}$ &1.13 & 7.18  \\	
    		  &$2.26$& 2 & 8 & 1.2 &957 &1.36  &n/a	  & $1.35 \times 10^{-8}$ &$5.53 \times 10^{-5}$ & 2.14 & 7.59  \\		      
\enddata
%\tablenotetext{a}{Type Ia abundances used in this study ($log (\textrm{X/H}) +12$):}  
\tablecomments{Type Ia abundances used in this study ($log (\textrm{X/H}) +12$)- He: 10.93, O: 12.69, C: 0.0, Ne:12.65, N: 0.0, Mg: 11.96, Si: 12.87, Fe: 13.13, S: 12.52, Ca: 11.97, Ni: 11.85, Na: 6.24, Al: 6.45, Ar: 11.81.}  
\end{deluxetable}

\clearpage

\begin{deluxetable}{ccrrrrrrr}
\tabletypesize{\scriptsize}
%\rotate
\tablecaption{Dimensionless Emission Measures and Temperatures for Self-similar Phases \label{tab:tabss}}%$^{a}$
\tablewidth{0pt}
\tablehead{
\colhead{phase}  & \colhead{s} & \colhead{n}  & \colhead{$C/\tau$ } & \colhead {$dEM_{FS}$} &  \colhead{$dT_{FS}$ } &  \colhead{$dEM_{RS}$ }
  & \colhead{$dT_{RS}$ }  
}						
\startdata
 WL91       &  n/a   &  n/a   &  0 &   0.5164 & 1.2896 &  n/a &  n/a \\  
         &  n/a  &  n/a  & 1 &  0.7741 & 1.3703 &  n/a &  n/a \\  
          &  n/a   &  n/a  & 2 &  1.6088 & 1.3693 &  n/a &  n/a \\
        &  n/a   &  n/a  & 4 &  6.9322 & 1.3833 &  n/a &  n/a \\
 CP    &  0  & 6   &  n/a &  0.6746 & 1.2614 & 0.1604 & 0.8868 \\  
  &  0  & 7   &  n/a &   0.7542 & 1.2227 &  0.6250 & 0.5204 \\  
    &  0  & 8   &  n/a &   0.8081 & 1.1922 & 1.549  & 0.3368 \\  
     &  0  & 9   &  n/a &   0.8471 & 1.1694 &  3.087 & 0.2347 \\  
     &  0   & 10   &  n/a &   0.8767 & 1.1518 &  5.400 & 0.1723 \\  
      &  0   & 11  &  n/a &   0.8998 & 1.1380 & 8.634 & 0.1318 \\  
      &  0    & 12  &  n/a &   0.9184 & 1.1270 &  12.98 & 0.1039 \\  
      &  0     & 13  &  n/a &   0.9337 & 1.1179 & 18.53  & 0.0840 \\  
      &  0    & 14  &  n/a &   0.9465 & 1.1103 & 25.50  & 0.0693 \\  
         CP    & 2  & 6   &  n/a &  17.60 & 0.1417 & 12.95 & 0.0413 \\  
  &  2  & 7   &  n/a &   99.35 & 0.0298 & 47.88 & 0.0254 \\  
    &  2  & 8   &  n/a &   56.92 & 0.0533 & 116.0  & 0.0169 \\  
     &  2  & 9   &  n/a &  45.84 & 0.0677 &  227.3 & 0.0119 \\  
     & 2   & 10   &  n/a &   62.73 & 0.0515 &  391.5 & 0.00887 \\  
      & 2  & 11  &  n/a &   79.16 & 0.0420 & 619.3 & 0.00684 \\  
      &  2    & 12  &  n/a &  94.66 & 0.0360 &  920.8 & 0.00644 \\  
      &  2    & 13  &  n/a &   75.18 & 0.0451 & 1308  & 0.00442 \\  
      & 2   & 14  &  n/a &   83.76 & 0.0411 & 1788  & 0.00367 \\   
\enddata
%\tablenotetext{a}{All models have SN ejecta mass of 1.4$M_{\odot}$. }
\end{deluxetable}

\end{document}